\documentclass[trackchanges, twocolumn]{aastex701}

\definecolor{changegreen}{RGB}{0,110,0}

\usepackage{amsmath}

\begin{document}

\title{A Multi-Mission Detection Framework for High-Energy Transients: Application to a 13-year Fermi/GBM and Swift/BAT Sample}

\author[0009-0004-8898-248X]{Ariel Perera}
\affiliation{Department of Particle Physics \& Astrophysics, Weizmann Institute of Science, Rehovot 76100, Israel}
\email[show]{ariel.perera@weizmann.ac.il}

\author[0000-0001-5162-9501]{Barak Zackay}
\affiliation{Department of Particle Physics \& Astrophysics, Weizmann Institute of Science, Rehovot 76100, Israel}
\email{barak.zackay@weizmann.ac.il}

\author[0000-0002-1661-2138]{Tejaswi Venumadhav}
\affiliation{Department of Physics, University of California at Santa Barbara, Santa Barbara, CA 93106, USA}
\affiliation{International Centre for Theoretical Sciences, Tata Institute of Fundamental Research, Bangalore 560089, India}
\email{teja@ucsb.edu}

\begin{abstract}

    The detection of faint high-energy transients is becoming increasingly important in time-domain and multi-messenger astronomy, particularly for identifying electromagnetic counterparts to binary neutron star mergers, high-energy neutrino events, and other weak transient phenomena. At present, the sensitivity of gamma-ray and hard X-ray searches is often limited by the detection threshold of individual instruments, leaving a population of faint transients inaccessible to single-mission analyses. To address this limitation, we develop a joint detection framework that combines trigger information from multiple detectors and assigns each candidate a joint association probability, $p_{\text{joint}}$, as a function of the individual detector signal-to-noise ratios. We apply this framework to archival Fermi/GBM and Swift/BAT rates data from 2013-2025. The method effectively separates coincident signals from accidental background associations, suppressing inconsistent coincidences while enhancing the significance of faint events observed by both instruments. We find that, under optimal conditions, a two-mission search can increase the accessible volume for short gamma-ray bursts by up to $60\%$. This framework provides a practical foundation for generalized multi-mission searches, enabling more sensitive exploration of the faint high-energy transient sky and improving the prospects for future multi-messenger discoveries.
        
\end{abstract}


\section{Introduction} \label{sec: introduction}
    The detection of weak gamma-ray and hard X-ray transients is becoming increasingly important with the rapid growth of time-domain and multi-messenger astronomy. The multi-messenger event GW170817 and its electromagnetic counterpart, GRB 170817A \citep{abbott_gravitational_2017, goldstein_ordinary_2017, savchenko_integral_2017}, demonstrated the scientific value of identifying faint gamma-ray transients. To date, GW170817 remains the only binary neutron star merger observed with an electromagnetic counterpart, among the only two binary neutron star mergers observed to date \citep{the_ligo_scientific_collaboration_gwtc-50_2026}. 
    
    As the LIGO-Virgo-KAGRA (LVK) network improves its sensitivity in preparation for the upcoming O5 observing run, the gravitational-wave detection horizon will continue to expand. Consequently, the associated electromagnetic counterparts are expected to become increasingly faint. Beyond gravitational-wave follow-up, optical transients such as kilonovae may also be detected independently by the Vera C. Rubin Observatory/LSST \citep{ivezic_lsst_2019,shvartzvald_ultrasat_2024}. Kilonovae are expected to be associated with compact binary mergers and short gamma-ray bursts, as demonstrated by GW170817, GRB 170817A, and AT2017gfo \citep{eichler_nucleosynthesis_1989,  abbott_gravitational_2017,coulter_swope_2017, smartt_kilonova_2017}. In addition, high-energy neutrinos may also be temporally and spatially coincident with GRBs (\cite{the_icecube_collaboration_multimessenger_2018}). Such joint detections can provide crucial information for constraining progenitor models and reveal the physical processes that power these emissions. In addition, lowering the detection threshold can also uncover faint or otherwise sub-threshold bursting activity from persistent sources, including soft gamma repeaters (SGRs) \citep{kaspi_magnetars_2017}, as well as rare pulsars that show transitional behavior between high-magnetic-field radio pulsars and magnetars \citep{blumer_psr_2017}.
    
    Fortunately, most space-based gamma-ray and hard X-ray surveys possess large fields of view. Even though a single observatory in low Earth orbit (LEO) is inevitably insensitive to approximately a third of the sky at any given moment due to Earth occultation, the concurrent operation of multiple missions can provide a nearly uniform coverage. Crucially, when the fields of view of different instruments overlap, combining their data is the most effective strategy for detecting increasingly faint events that would otherwise remain hidden in the background noise of any single detector.
    
    Historically, collaborative efforts between spatially dispersed observatories have been highly successful in improving the localization of gamma-ray bursts, most notably through the InterPlanetary Network (IPN) \citep{hurley_interplanetary_2013}. By utilizing the large interplanetary baselines between spacecraft—such as the Mars Odyssey orbiter, Konus-Wind at the L1 Lagrange point, and Fermi/GBM in LEO—the IPN localized bursts by triangulations, achieving localizations ranging from a few square arcminutes to several square degrees, depending on burst brightness and spacecraft geometry \cite{svinkin_second_2022}. However, while the IPN showcases the strength of combining cross-instrument data, it primarily relies on timing relatively bright, already-triggered events. 
    
    To dig deeper into the noise, targeted sub-threshold searches have been developed. These typically exploit coherence and utilize temporal templates within externally motivated time windows \citep[e.g.,][]{kocevski_analysis_2018}. Yet, there remains a need for generalized frameworks that evaluate the joint probability of blind, un-targeted sub-threshold signals across multiple instruments. A wealth of currently operating missions can be leveraged for such an analysis, including Fermi/GBM, Swift/BAT, GECAM, CALET, Glowbug, Konus-Wind, SVOM, MAXI, and HXMT-Insight, alongside small cubesat missions like GRBAlpha, GRBBeta and BlackCat \citep{meegan_fermi_2009, barthelmy_burst_2005, li_gecam_2020, li_inflight_2021, yamaoka_calet_2013, grove_glowbug_2020, aptekar_konus-w_1995, wei_deep_2016, matsuoka_maxi_2009, zhang_overview_2020, pal_grbalpha_2023, ripa_grbalpha_2026, chattopadhyay_blackcat_2018}. 
    
    In this work, we expand upon the joint-analysis framework developed in \cite{perera_expanding_2026}, which utilized Swift/BAT rates to validate Fermi/GBM triggers, including triggers well below the GBM detection threshold. By treating the overlapping observations not merely as independent verifications, but as components of a two-dimensional joint probability ($p_{\rm joint}$), we aim to extract faint events that would not have crossed the independent detection threshold of either instrument, finding that under optimal alignment of the two instruments, the available detection volume can increase by up to 60\%. The methodology presented here serves as a blueprint for combining diverse operating high-energy missions. 
    
    This paper is organized as follows: In Section \ref{sec: data acquisition}, we outline the data selection process and summarize the independent GBM and BAT pipelines. In Section \ref{sec: method}, we detail the formulation of the joint detection probability and the statistical methods used to construct the smoothed two-dimensional background and signal models. Section \ref{sec: results} presents the resulting joint probability distribution, discusses background contamination, and quantifies the fractional increase in sensitive detection volume. Finally, we summarize our findings and discuss the broader implications for future high-energy mission networks in Section \ref{sec: summary}.

\section{Data Selection} \label{sec: data acquisition}
    The analysis presented in this work builds upon two previously developed search pipelines. The first is the GBM transient detection pipeline described in \cite{perera_new_2025}, which was used to search the full 2013-2025 Fermi-GBM TTE data set. The second is the BAT rates follow-up analysis described in \cite{perera_expanding_2026}, which evaluates the presence of temporally coincident signals in Swift-BAT for candidate events identified by the GBM search.
    
    Here we give a brief summary of the results presented in these papers that are used in this work. The purpose of this section is to describe the origin of the trigger samples and the detection statistics used to construct the joint probability distribution.
    
    \subsection{GBM Detection Pipeline}
        Our GBM GRB detection pipeline \citep{perera_new_2025} operates on continuously transmitted time-tagged-event (TTE) data. Utilizing all 14 GBM detectors and their available energy channels, the pipeline applies a coherent Poisson matched-filter detection statistic. We employ stochastically placed spectral templates, derived from the GBM response at a fixed time and folded through a comprehensive sample of Band function \citep{band_batse_1993} spectra. The temporal templates consist of simple boxcars distributed logarithmically to ensure a maximum signal-to-noise ratio (SNR) loss of $7\%$ across adjacent templates. The longest template is $6.573$~s; for long bursts, this search timescale is best interpreted as a peak-flux duration rather than a full burst duration such as $T_{90}$. Longer bursts than this duration are also recovered and assigned this duration by our pipeline. This was explored in our GRB catalog paper \citep{perera_expanding_2026}.
        
        To account for non-stationary and non-Gaussian noise, the pipeline incorporates a background drift correction \citep[defined in][]{perera_new_2025}, conceptually similar to the power spectral density (PSD) drift corrections utilized in gravitational-wave data analysis \citep{zackay_detecting_2021}. We convert each trigger SNR into equivalent Gaussian standard deviations by calculating the $p$-value and mapping it to the standard normal distribution. This calibration is performed per temporal and spectral template, relying on Monte Carlo simulations of the template's null distribution.
        
        Applying this pipeline to the complete 13-year archive of GBM TTE data (2013-2025) yielded a comprehensive catalog of triggers. Each trigger subsequently underwent parameter estimation and a classification scheme that categorizes the source based on its inferred properties.

        Because storing the complete SNR time series for the full 13-year GBM
        archive is prohibitively expensive, triggers were recorded down to
        ${\rm SNR} \geq 5.1$,
        far below the single-instrument detection threshold
        \citep{perera_expanding_2026}. The BAT follow-up of these deeply
        sub-threshold triggers therefore probes coincidences well below the GBM
        detection limit, although the search remains seeded by GBM: a transient
        below this recording floor, however bright in BAT, is not recoverable by
        the present analysis.

    \subsection{BAT Rates Follow-up Analysis} \label{sec: bat rates}
        For every trigger identified during the GBM detection and parameter estimation phase, we conducted a targeted search for a coincident signal within the Swift/BAT rate data. Because BAT continuously provides integrated detector-plane data binned at 0.064 s, 1.0 s, and 1.6 s resolutions, we leverage the GBM trigger times to search for transient rate excesses that are temporally aligned (a one bin timing offset was permitted to account for the combined effects of data binning, light-travel time between the spacecraft, and any clock uncertainties) and have durations consistent with the GBM event.
        
        Using the detection statistic developed in \cite{perera_expanding_2026}, we evaluated these temporally coincident windows alongside shifted, unassociated time windows to characterize the SNR distribution of false associations. We denote these values as "ontime" and "shifted," respectively, throughout the paper. To obtain the shifted-trigger distribution, we applied exactly the same procedure used for the ontime triggers, but shifted the GBM trigger time by 250 sec for each trigger. This shift was chosen to preserve similar BAT background characteristics to those in the ontime windows while increasing the likelihood that BAT data are available at the shifted time, for example when Swift has not yet slewed to the GBM trigger position. As with the GBM pipeline, we applied a drift correction to each candidate to mitigate noise arising from non-stationary background variations. This procedure ultimately yields a distribution of cross-matched SNRs (Fig. \ref{fig: raw snrs}). 
        
    \subsection{Selection of Coincident Triggers}
\begin{figure}
            \centering
            \includegraphics[width=1.0\linewidth]{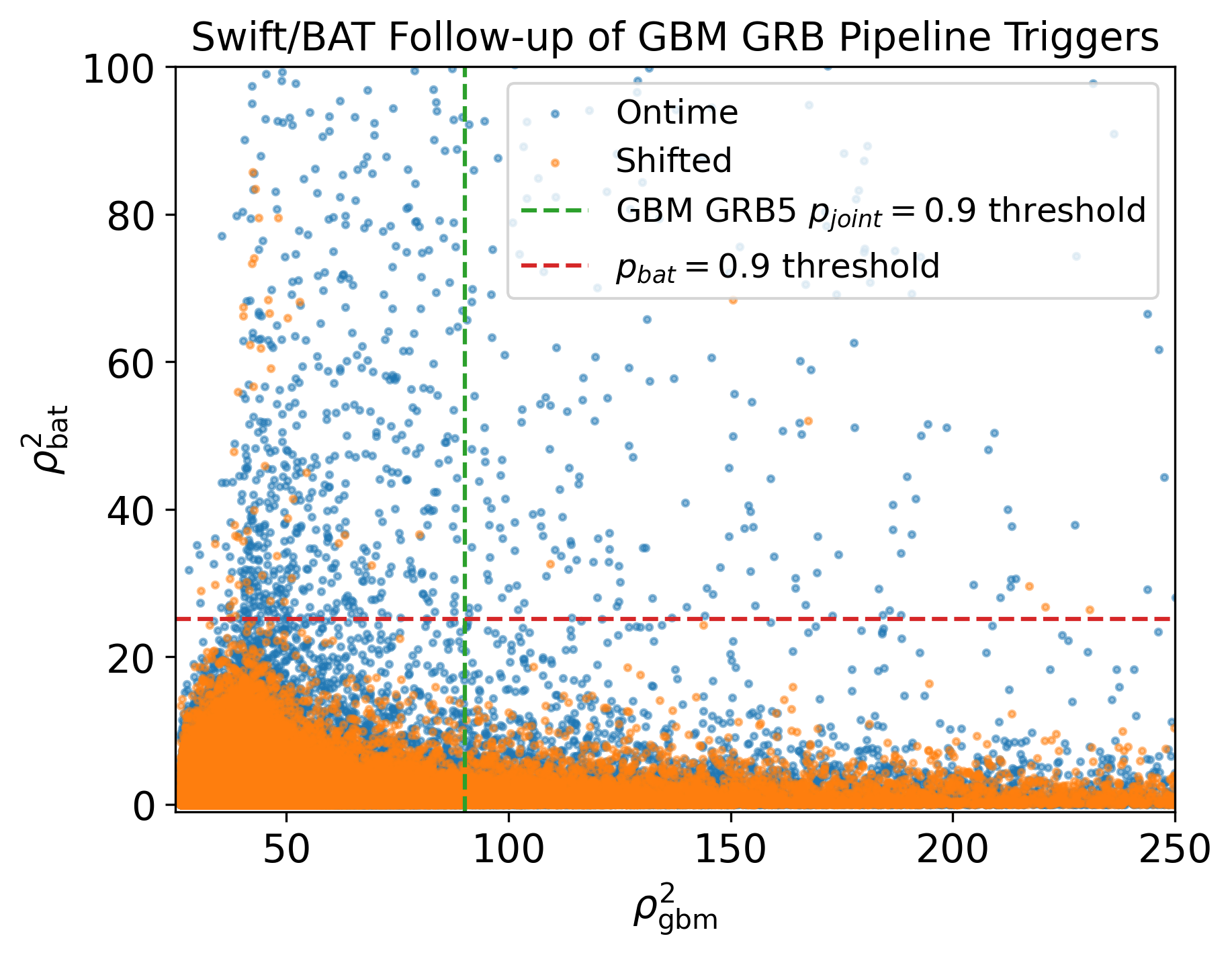}
            \caption{\textbf{2D SNR${}^2$ distribution from the GBM and BAT detectors.} $\rho^2_{\text{gbm}}$ and $\rho^2_{\text {bat}}$ are the SNR${}^2$ values obtained from GBM and BAT respectively. Blue points correspond to $\rho^2_{\text {bat}}$ at the exact time of the GBM trigger, while orange points represent the $\rho^2_{\text {bat}}$ at shifted, non-associated times (background). The dashed lines outline the individual detector thresholds. The red dashed line indicates the BAT threshold at $p_{\text{bat}}=0.9$ (agnostic to trigger classification), and the green dashed line corresponds to the GBM threshold for the GRB5 class at $p_{\rm astro}=0.9$. Because thresholds vary across different GRB classes, the GRB5 class is used here for illustration. Regions extending beyond these green and red lines represent high-probability triggers. Notably, an excess of ontime triggers is visible even at values below the individual detector thresholds, motivating the joint analysis.}
            \label{fig: raw snrs}
        \end{figure}

        All triggers identified by the GBM detection pipeline were followed up with Swift/BAT whenever rate data around the trigger time were available. For the present analysis, we retain only triggers classified as GRB, SGR, or ambiguous GRB/SGR, following the scheme of \cite{perera_expanding_2026}. The resulting two-dimensional distribution of SNRs from both detectors is shown in Fig.~\ref{fig: raw snrs}.

        For the joint analysis detailed in the following section, we restrict our sample to triggers with $\rho^2_{\text{gbm}} \leq 150$ and $\rho^2_{\text{bat}} \leq 60$. Events exceeding these SNR ratios are highly significant detections in the corresponding instrument, but their probability of association is not necessarily close to unity when the SNR in the other instrument remains low. After deriving the joint probability, $p_{\text{joint}}$, within this bounded region, we evaluate higher-SNR events using nearest-neighbor extrapolation from the boundary of the sampled region. This extrapolation preserves the dependence of the association probability on the SNR in both instruments, such that events with high $\rho^2_{\text{gbm}}$ but low $\rho^2_{\text{bat}}$ can still have relatively low $p_{\text{joint}}$.

\section{Method} \label{sec: method}
    To quantify whether a GBM trigger has a counterpart in the BAT rate data, we extend the $p_{\rm astro}$ and $p_{\rm bat}$ framework of \cite{perera_expanding_2026} to the plane $x=(\rho^2_{\rm gbm},\rho^2_{\rm bat})$. We consider two hypotheses for an on-time pair: $s$, the GBM trigger and the BAT excess arise from the same astrophysical transient; and $n$, they do not. The latter includes pairs in which the GBM trigger is astrophysical but the transient is not detectable in the BAT rates (e.g., outside the coded field of view). We define the joint association probability as the posterior probability of $s$
    \begin{equation} \label{eq: pjoint}
        p_{\text{joint}} = \frac{p(\rho^2_{\text{gbm}}, \rho^2_{\text{bat}}|s ) p(s)}
        {p(\rho^2_{\text{gbm}}, \rho^2_{\text{bat}} | s)p(s) + p( \rho^2_{\text{gbm}}, \rho^2_{\text{bat}} | n)p(n)},
    \end{equation}
    with $p(s)=1-p(n)$. Equivalently, the posterior odds are $p_{\rm joint}/(1-p_{\rm joint}) = \mathcal{B}(x)\,p(s)/p(n)$, where $\mathcal{B}(x)=p(x|s)/p(x|n)$ is the Bayes factor. $p_{\rm joint}$ is thus not a detection efficiency: it does not quantify how often a transient of given brightness is recovered, but the relative credence of the association hypothesis for a pair at $x$. Its ranking of candidates is set by $\mathcal{B}$; its absolute calibration additionally depends on the relative rates of genuine and accidental coincidences in this sample. Because a genuine association requires an astrophysical GBM trigger, $p_{\rm joint}$ is a lower bound on the probability that the GBM trigger is astrophysical.
    
    In practice, the distribution of events obtained by matching the SNR values
    at the same time in both detectors, denoted the ontime distribution or
    $\mathcal{H}_1$, contains both true associations and accidental
    coincidences, while matching at randomly shifted times samples only the
    accidental-coincidence (noise) distribution, denoted $\mathcal{H}_0$ (see
    Sec.~\ref{sec: bat rates} for the construction of the ontime and shifted
    samples). Modeling the ontime sample as a two-component mixture of signal
    and noise, Eq.~(\ref{eq: pjoint}) can be written entirely in terms of the
    two measured number densities (see Appendix~\ref{appendix: pastro} for the
    derivation):
    \begin{equation} \label{eq: pjoint empirical}
        p_{\text{joint}} =
        \frac{n_{\rm ontime}(\rho^2_{\rm gbm}, \rho^2_{\rm bat})
              - \alpha\, n_{\rm shifted}(\rho^2_{\rm gbm}, \rho^2_{\rm bat})}
             {n_{\rm ontime}(\rho^2_{\rm gbm}, \rho^2_{\rm bat})},
    \end{equation}
    where $n_{\rm ontime}$ and $n_{\rm shifted}$ are the trigger number
    densities of the ontime and shifted samples, and
    $\alpha = (1-f)\,N_1/N_0$ accounts for the different effective sizes of
    the two samples, with $f$ the signal fraction of the ontime sample and
    $N_1$, $N_0$ the total ontime and shifted counts.
    
    \begin{figure*}
        \centering
        \includegraphics[width=1.0\linewidth]{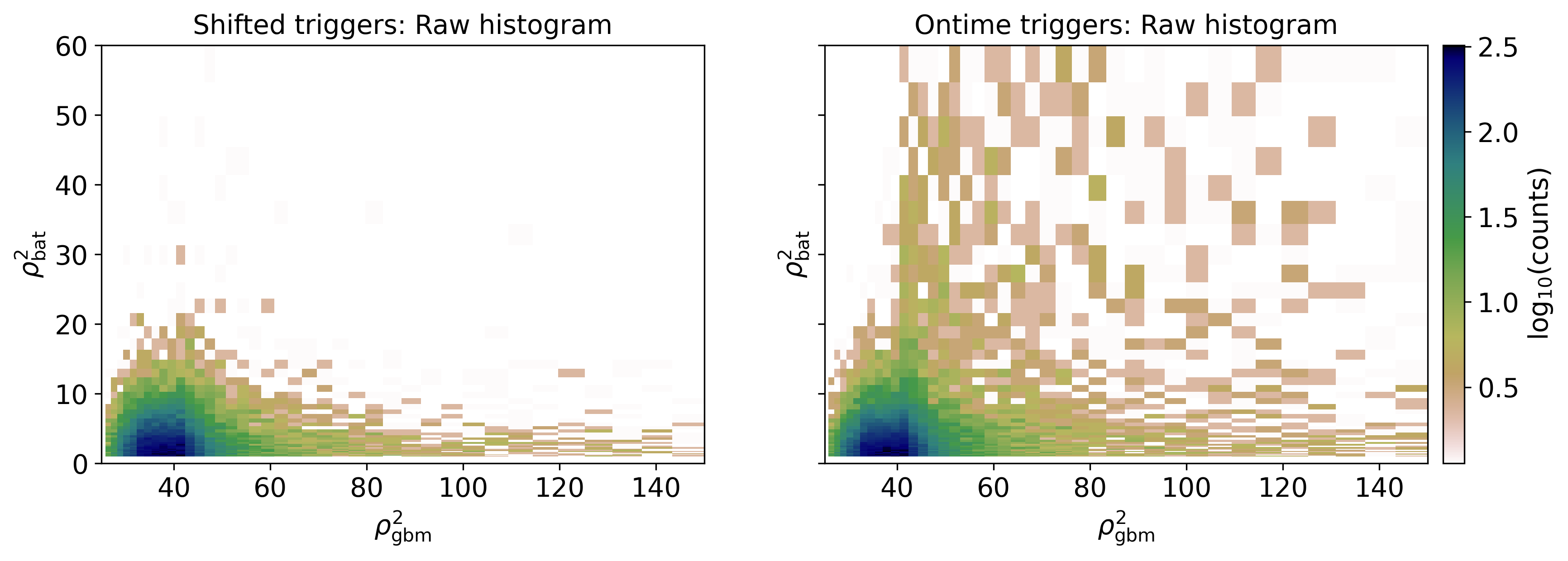}
        \includegraphics[width=1.0\linewidth]{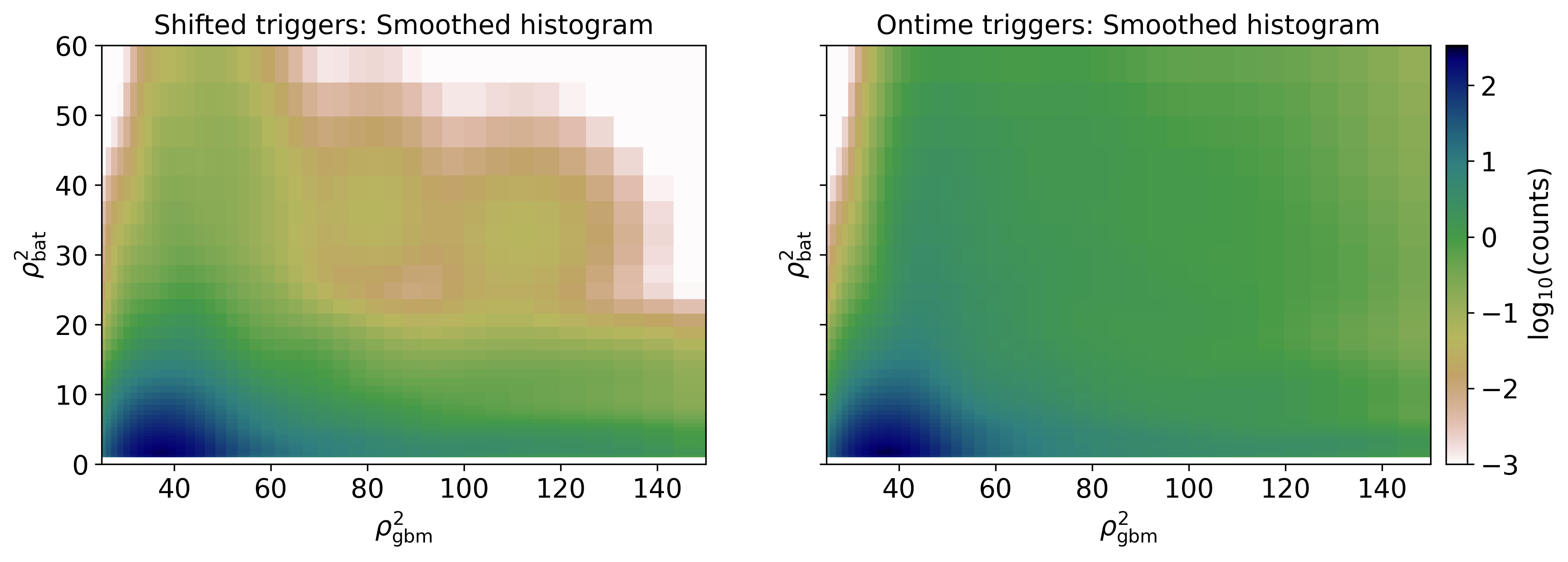}
        \caption{\textbf{Raw and smoothed trigger histograms}. Top panel: The raw histograms that are obtained by using logarithmic bins in each axis. The trigger density of the ontime triggers is seen to be much higher in the higher SNR regime due to the overdensity of triggers originating from actual signals. Nevertheless, the many empty bins in both ontime and shifted triggers require a smoothing of the histogram for the calculation of the joint detection probability. Bottom panel: The smoothed histograms that are obtained by applying a Gaussian filter to the 2D histograms in the upper panel.}
        \label{fig: 2d hists}
    \end{figure*}
    As shown in Appendix~\ref{appendix: pastro}, $\alpha$ equals the ratio of
    ontime to shifted counts in any region of the plane that is free of true
    associations. We estimate it from the background-dominated region
    $|\rho_{\rm gbm}^2 - 35| < 2$, $|\rho_{\rm bat}^2 - 5| < 2$, which contains ontime and shifted triggers, giving
    $\alpha = 1.13$, from $N_1=3303$ and $N_0 = 2925$ counts in the ontime and shifted sample respectively. The full ontime and shifted samples contain $N_1=49,504$ and $N_0=40,536$ pairs, respectively, giving $N_1/N_0=1.22$. Thus, the measured $\alpha=1.13$ is consistent with the mixture model and corresponds to an inferred association fraction $f=0.075$.
        
    Evaluating $p_{\rm joint}$ over the full $(\rho_{\rm gbm}^2,\rho_{\rm bat}^2)$ plane requires smooth estimates of the underlying probability density functions. This is challenging because trigger rates decrease rapidly with increasing SNR (Fig. \ref{fig: raw snrs}), resulting in sparse sampling and large statistical fluctuations in the high-SNR regime. A direct estimate based on a two-dimensional histogram, therefore, contains many empty or poorly populated bins.
    
    To mitigate these fluctuations, we smooth the histograms with a Gaussian kernel and interpolate the resulting densities using bivariate splines. While kernel smoothing introduces bias by broadening sharp density gradients, this effect is most pronounced in sparsely populated regions where the inferred probabilities are already close to unity due to the high significance of the triggers. Near the signal-noise transition region, where the trigger density remains relatively high and $p_{\rm joint}$ is most sensitive to the density estimate, the impact of smoothing is comparatively small. This motivates the following procedure for estimating $p_{\rm joint}$:
    
    \begin{itemize}
        \item Construct two-dimensional histograms of $(\rho_{\rm gbm}^2,\rho_{\rm bat}^2)$ for the ontime and shifted trigger populations using logarithmic binning along both axes.
        \item Apply a Gaussian smoothing kernel to each histogram.
        \item Fit a bivariate cubic spline to the smoothed histograms
        \item Evaluate the spline representations on a dense two-dimensional grid.
        \item Compute $p_{\rm joint}$ using Eq. \ref{eq: pjoint empirical} at each grid point, clipping to [0,1] where $\alpha n_{\text{shifted}}>n_{\text{ontime}}$.
        \item For an arbitrary trigger, obtain $p_{\rm joint}$ by interpolating within the resulting probability grid.
    \end{itemize}
    
    The raw and smoothed histograms are shown in Fig. \ref{fig: 2d hists}. The ontime distribution exhibits a clear excess relative to the shifted distribution at high SNRs, indicating the presence of a substantial signal population. The numerous empty bins visible in the raw histograms are primarily a consequence of finite sampling rather than a true absence of events, further motivating the use of smoothing and interpolation when estimating the joint detection probability.

    In Appendix~\ref{app: robustness}, we present robustness tests in which we vary the kernel width and bootstrap the triggers. The bootstrap procedure also provides uncertainty estimates for the quantities reported later in the paper and extracted using the procedure described above.

\section{Results and discussion} \label{sec: results}
    \subsection{The joint probability distribution}
\begin{figure*}
            \centering
            \includegraphics[width=1\linewidth]{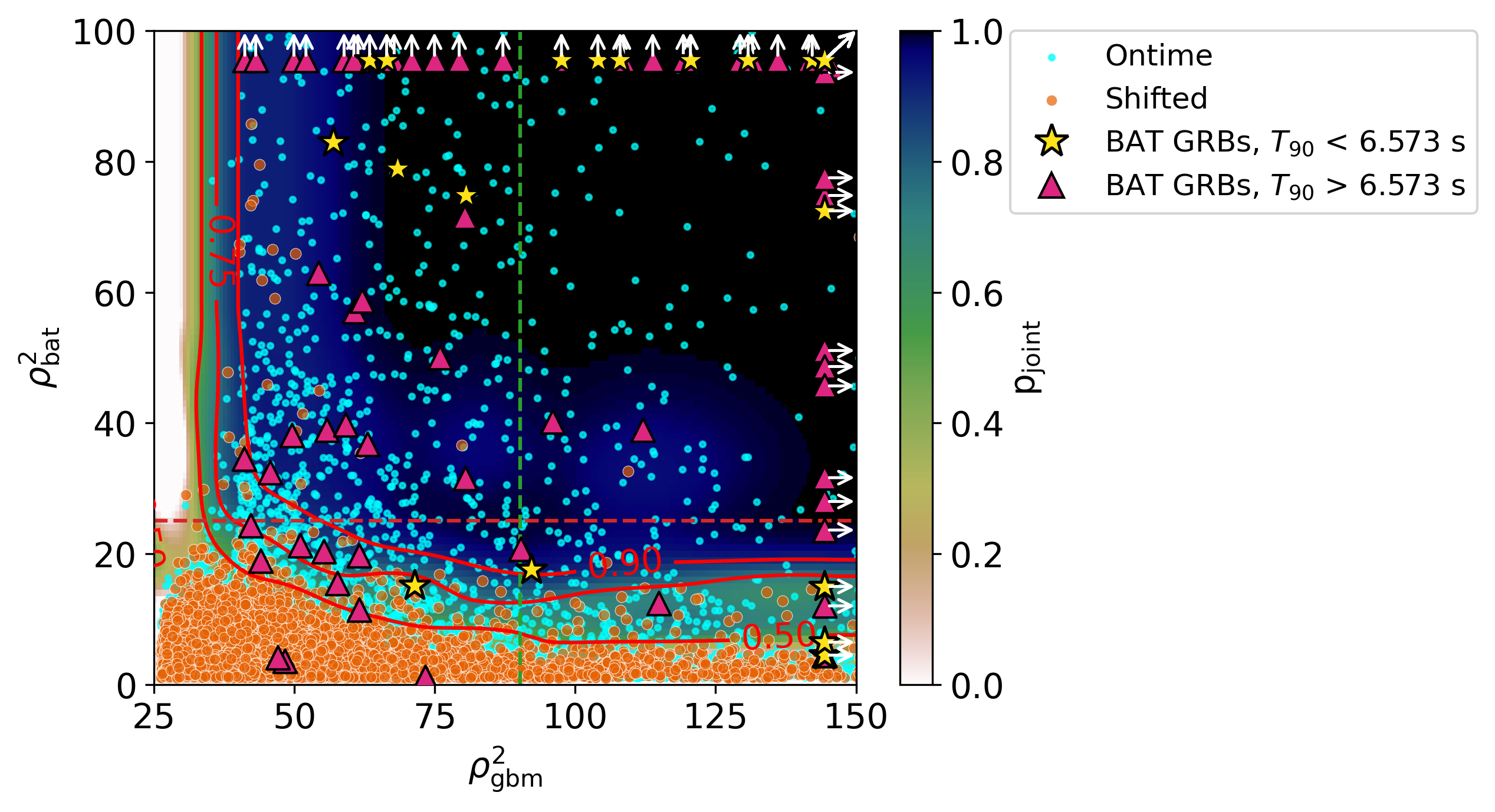}
            \caption{\textbf{$p_{\text{joint}}$ distribution}. The colormap shows the 2D $p_{\text{joint}}$ distribution, overlaid with the individual triggers that contributed both to the ontime distribution and to the shifted distribution. Contours of 0.5, 0.75 and 0.9 are shown to guide the eye, as well as the threshold for 0.9 probability of detection in BAT and GRB5 in GBM (same as in Fig. \ref{fig: raw snrs}). The $p_{\text{joint}}=0.9$ contour extends below the intersection of the individual-instrument $0.9$ thresholds (red and green dashed lines), enlarging the region of parameter space in which coincident triggers qualify as high-confidence detections. The white stars and triangles mark triggers that correspond to BAT catalog GRBs, where we separated the associations to bursts with $T_{90}$ larger or smaller than 6.573 seconds, which corresponds to our maximal temporal template in the search pipeline. Bursts lying beyond the limits of the figure are indicated by white arrows pointing in the direction of increasing detector SNR.}
            \label{fig: pjoint}
        \end{figure*}

        The main result of this paper is the $p_{\text{joint}}$ distribution that was calculated using Eq. \ref{eq: pjoint} and is presented visually in Fig. \ref{fig: pjoint}.
        The figure shows the inferred joint association probability between GBM and BAT as a function of the SNR${}^2$ in the two instruments. Several features are immediately apparent. First, the probability increases with increasing significance in both detectors, as expected for a population dominated by real astrophysical events. Second, the probability contours are not aligned with either the GBM or BAT axes, indicating that the significance measured by both instruments contributes to the final probability estimation. 
        
        The $p_{\rm joint}=0.9$ contour extends into regions of parameter space where neither instrument alone would necessarily assign $p=0.9$. This demonstrates the main advantage of a coincidence-based search. Triggers appearing in two independent instruments can together boost the significance of an event even when the significance provided by either detector individually is lower. 

    \subsection{Recovery of Swift/BAT catalog GRBs} \label{sec: catalog recovery}
        As a validation of the joint statistic, we cross-matched our triggers against the Swift/BAT GRB catalog\footnote{\texttt{https://swift.gsfc.nasa.gov/results/batgrbcat/}} (updated through 2025-06-05). A trigger is associated with a catalog GRB when its trigger time falls within the burst's $T_{90}$ interval, padded by 1~s on each side, and its GBM maximum likelihood estimated (MLE) sky localization lies within $30^\circ$ of the BAT catalog reported position - an angular offset found to provide a reasonable matching criterion in the parameter estimation study of \cite{perera_expanding_2026}. The positional requirement rejects chance temporal matches with inconsistent sky locations.
        In total, 155 catalog GRBs are recovered among our triggers; the complete list is given in Table~\ref{appendix: bat recovered grbs} in the appendix. We also repeated the exact same matching procedure using shifted BAT catalog times, at offsets of $\pm1000$, $\pm2000$, and $\pm10,000$ sec. We find, on average, 2 false associations. This corresponds to $\sim1\%$ of the sample. The vast majority of the associations (135 of 155) are assigned $p_{\rm joint} \geq 0.9$, and most lie far beyond the axis ranges of Fig.~\ref{fig: pjoint}. All bursts are overlaid in the figure, with the ones that falls outside the limits are marked with arrows, and the 32 GRBs that fall within limits of the figure are marked in Table~\ref{appendix: bat recovered grbs}. The bursts are separated to $T_{90}$ smaller or larger than 6.573~s, the longest temporal template used in the search pipeline. Table \ref{appendix: bat recovered grbs} also lists the $T_{90}$ durations reported in the GBM catalog (the continuously updated Fermi/GBM Burst Catalog maintained by the Fermi Science Support Center) for bursts associated with cataloged GBM events. This column is left blank when no GBM association exists. We identify 29 GRBs with $p_{\text{joint}} > 0.9$ that are absent from the GBM catalog.

        Instructive validation cases are catalog GRBs that are sub-threshold in our detection pipeline in each instrument individually. GRB130327A ($\rho^2_{\rm gbm}=92.2$, $\rho^2_{\rm bat}=17.6$) was assigned $p_{\rm astro}=0.82$ and $p_{\rm bat}=0.69$, below the high-confidence threshold of either instrument alone, yet the joint statistic promotes it to $p_{\rm joint}=0.93$. Similarly, GRB160127A ($\rho^2_{\rm gbm}=71.5$, $\rho^2_{\rm bat}=15.2$) was assigned $p_{\rm astro}=0.58$ and $p_{\rm bat}=0.63$, while the joint statistic yields $p_{\rm joint}=0.69$, exceeding both. Both bursts have $T_{90} \simeq 6$~s, comparable to our longest temporal template, so their modest SNRs reflect genuine faintness at the searched timescales or highly variable background rather than a significant duration mismatch. The angular offsets between the pipeline localizations and the BAT-reported positions ($8.2^\circ$ and $5.8^\circ$, well within the localization uncertainty of faint GBM triggers) provide another confirmation that these associations are genuine. The corresponding rates data and sky localizations from both GBM and BAT are presented in Appendix \ref{appendix: gbm bat plots}. Catalog GRBs recovered in this regime provide evidence that the population promoted by the joint statistic is astrophysical.

        Conversely, a minority of catalog GRBs (20 of 155) receive $p_{\rm joint}<0.9$, and even very bright bursts can appear at surprisingly low SNRs. Both effects share a common origin: the pipeline is optimized for short transients. The matched filter integrates at most 6.573 s of a burst, limiting the recovered SNR for long-duration events. In addition, the background drift correction, which compensates for background misestimation, substantially reduces the recovered SNR because the slowly varying emission of long bursts is partially absorbed into the background estimate, resulting in an inaccurate background model. As a result, long catalog GRBs are often recovered not at their main emission episode but through a short emission period, such as brief peaks following or preceding the main event. GRB131024B and GRB151006A ($T_{90} = 97$~s and $211$~s) illustrate this behavior: both are bright BAT catalog bursts, yet the pipeline recovers only weak, short peaks ($\rho^2_{\rm gbm} = 45.7$ and $49.6$, matched at the $0.133$~s and $2.671$~s templates), which the joint statistic nevertheless correctly assigns $p_{\rm joint}>0.9$. Consistent with this picture, 16 of the 20 bursts with $p_{\rm joint}<0.9$ have BAT $T_{90} > 6.573$~s, and 27 of the 32 catalog GRBs falling within the plotted region of Fig.~\ref{fig: pjoint} are long bursts scattered toward low SNRs. A few short bursts are also recovered with low $\rho^2_{\rm bat}$, resulting from differences between the durations observed, the different duration metric used - peak-flux interval and $T_{\text{90}}$, and highly varying background. We emphasize that a low $p_{\rm joint}$ is a statement about the strength of the coincident signal in the rate data at the searched timescales, not evidence against the astrophysical nature of the catalog burst; BAT catalog detections additionally exploit image-domain information that is unavailable to the rates-based statistic.

    \subsection{Background contamination}
        The shifted population is expected to provide an empirical estimate of the accidental coincidence background. However, Fig. \ref{fig: pjoint} shows that a small number of shifted triggers populate regions of the plane associated with high astrophysical probability. For example, the shifted trigger located near $(\rho_{\rm gbm}^2 \simeq 110$, $\rho_{\rm bat}^2 \simeq 35)$. Such events are unlikely to be purely statistical fluctuations. Instead, they likely correspond to real astrophysical transients that were incorrectly paired during the shifting procedure, to detector glitches, or to other unmodeled phenomena.
        
        The presence of these events broadens the estimated background distribution and therefore reduces the inferred value of $p_{\rm joint}$. A similar effect was observed in the single-instrument BAT analysis of \cite{perera_expanding_2026}. Because the available rate data contain only coarse spectral and detector information, identifying and removing such events remains challenging. Nevertheless, improved event characterization and filtering strategies are expected to further increase the sensitivity of future joint searches.

    \subsection{Down-weighting of inconsistent associations}
\begin{figure*}
            \centering
            \includegraphics[width=0.45\linewidth]{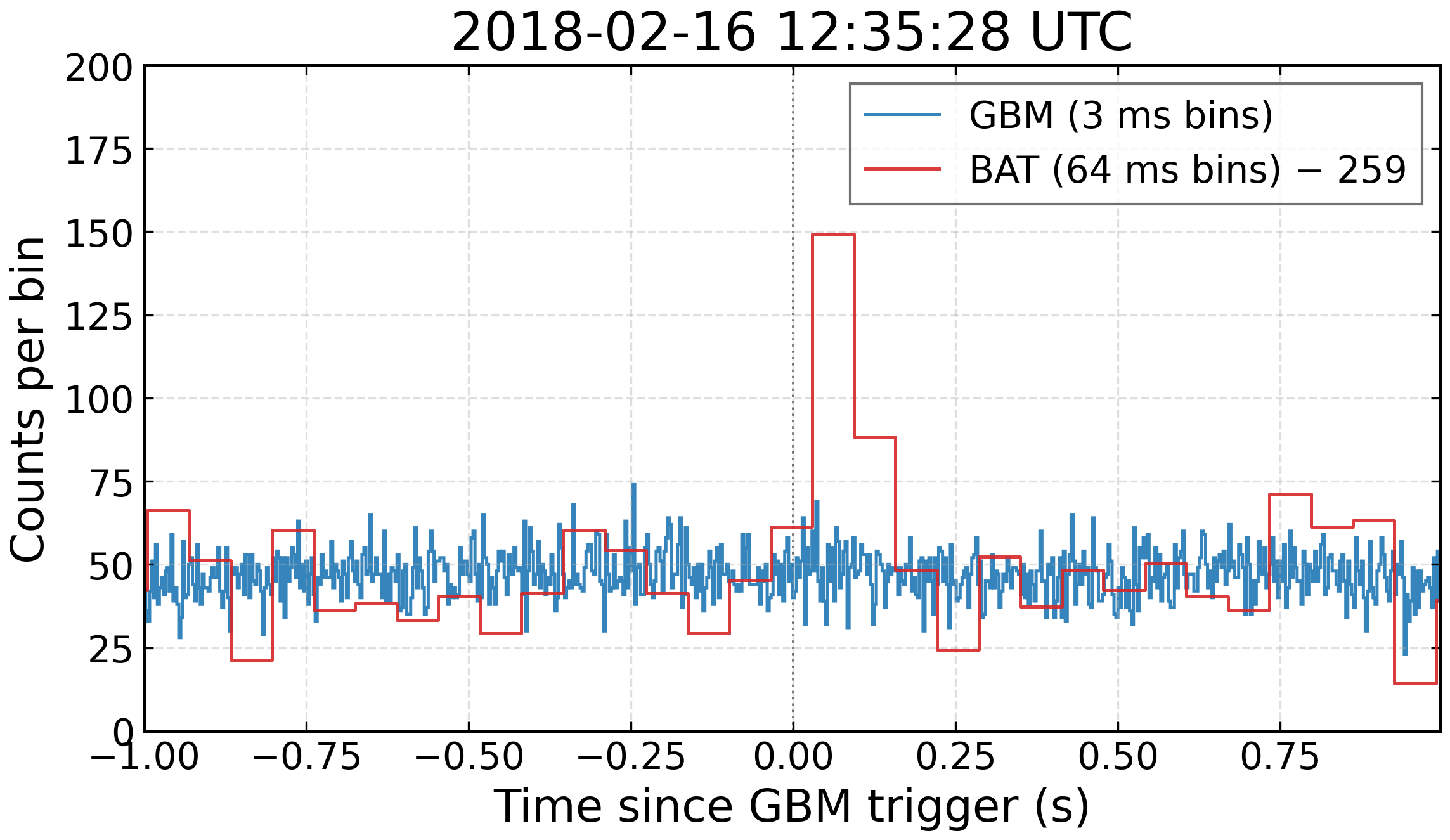}
            \includegraphics[width=0.5\linewidth]{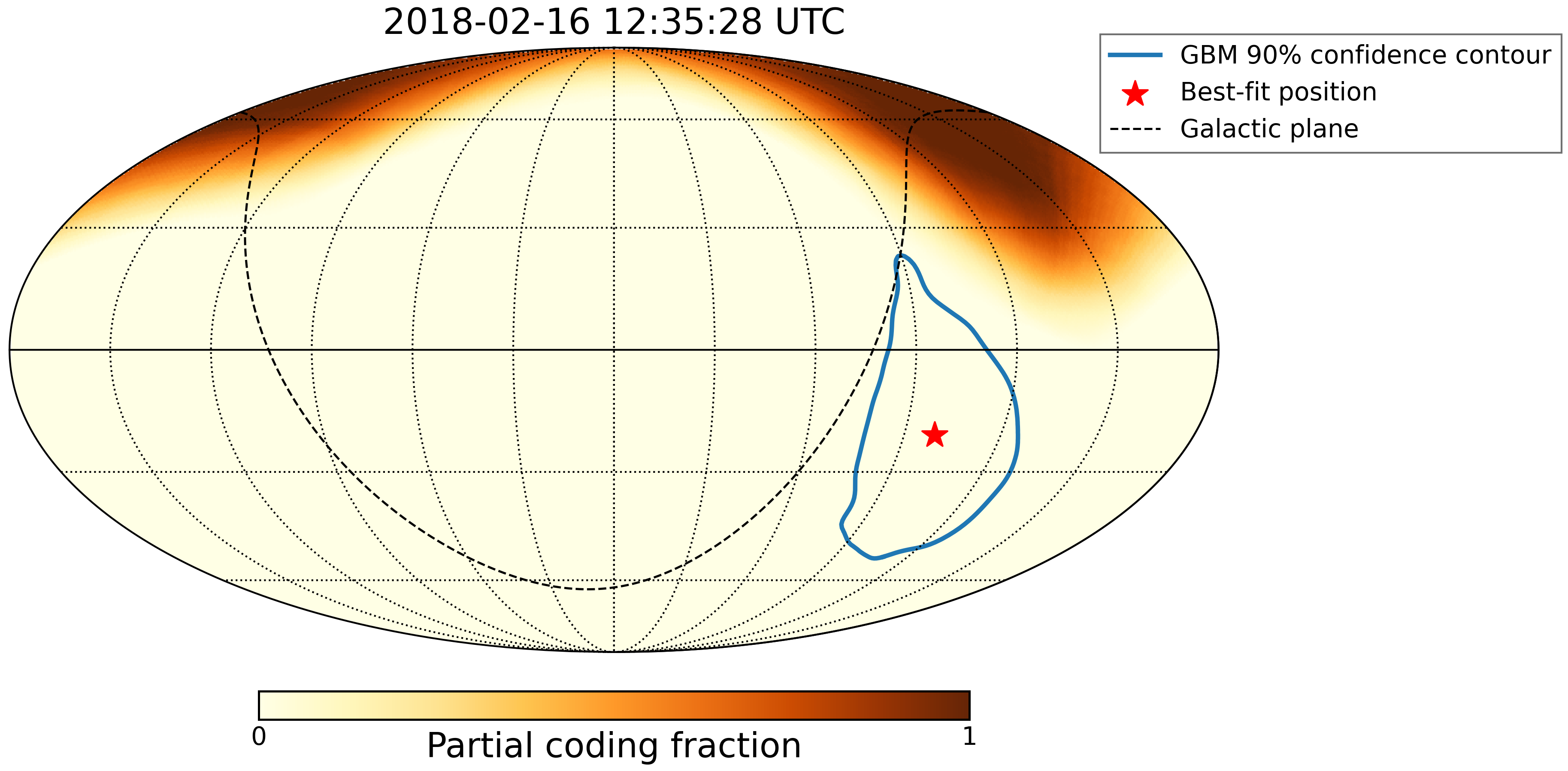}
            \caption{\textbf{Rates and sky localization of a trigger with $p_{\rm astro}=0$.}
            \textit{Left:} BAT rate data and GBM photon counts summed over their respective energy channels. The BAT detection statistic is computed from this summed light curve, whereas the GBM detection pipeline employs a two-dimensional matched filter using spectral templates across the energy channels, which are summed here for illustration only. The BAT rates have been shifted vertically to match the GBM background level for visual comparison. The highest-SNR GBM template has a duration of 3 ms, although the trigger remains consistent with $p_{\rm astro}=0$.
            \textit{Right:} The GBM localization is overlaid on the BAT partial-coding map. An excess is visible in the BAT light curve one time bin from the GBM trigger (the allowed offset), but the low significance of the GBM event results in a joint probability of only $p_{\rm joint}=0.13$. Furthermore, the GBM localization lies outside the BAT coded field of view. Although this does not rule out a common astrophysical origin, it provides additional evidence that the two triggers are unrelated, illustrating how the joint statistic appropriately downweights associations that lack support from both instruments.}
            \label{fig: short downweight}
        \end{figure*}

        An important feature of the joint statistic is its ability to suppress associations that are strongly supported by one instrument but not by the other. As an example, a very faint trigger occurring at 2018-02-16 12:35:28.187 UTC in GBM was assigned a significance consistent with $p_{\rm astro} = 0$, while the corresponding BAT trigger was assigned $p_{\rm bat}=0.93$. The resulting joint probability is only $p_{\rm joint}=0.13$. This behavior is desirable. While the BAT trigger is likely associated with a real astrophysical event, the GBM trigger is not necessarily astrophysical and is probably not from the same source. The joint statistic therefore correctly reduces the confidence of the association. A manual examination of the event supports this interpretation. The 90\% confidence localization region derived from GBM is inconsistent with the BAT field of view. Although this does not completely exclude a common origin, it provides additional evidence that the two triggers are unrelated. The corresponding rates and sky localizations are shown in Fig. \ref{fig: short downweight}.
        
    \subsection{Significance assessment and calibration}
\begin{figure}
        \centering
        \includegraphics[width=1\linewidth]{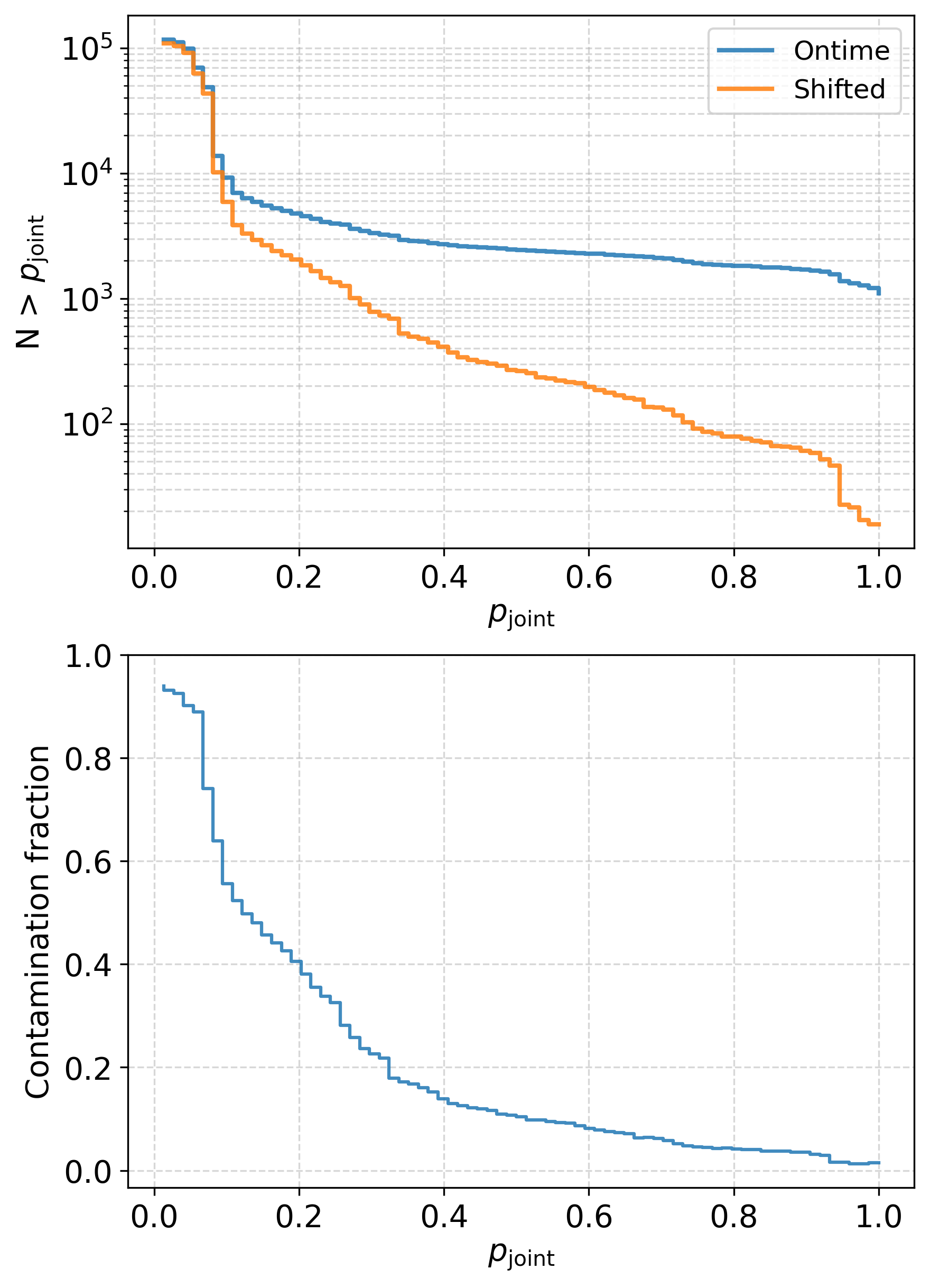}
        \caption{\textbf{Trigger contamination assessment.} Top: Cumulative number of triggers exceeding a given $p_{\text{joint}}$ threshold, evaluated for both the ontime and shifted distributions derived in Figure \ref{fig: pjoint}. Bottom: the ratio $\alpha N_{\text{shifted}}/N_{\text{ontime}}$ for triggers above a given $p_{\text{joint}}$, i.e., the estimated contamination fraction. The contamination drops rapidly as $p_{\text{joint}}$ increases, falling to $4\%$ at $p_{\text{joint}}=0.9$.}
        \label{fig: contamination}
        \end{figure}

        To calibrate $p_{\text{joint}}$ and account for high-SNR contamination from unrelated events, we calculate the cumulative number of candidates above a given probability threshold for both the ontime and shifted distributions. The bottom panel of Figure \ref{fig: contamination} provides a calibration curve and establishes the contamination fraction. We also calculate the false alarm rate (FAR) by dividing the cumulative number of background triggers by the observation time, FAR $= \alpha N_{\rm shifted}(p_{\rm joint}>p)/T_{\rm obs}$. We take $T_{\rm obs}$ to be the GBM trigger catalog time span multiplied by 0.65 to account for Earth occultation and the South Atlantic Anomaly passage (SAA). Table \ref{tab: joint numbers} details the trigger excesses, the resulting contamination fractions, and the FAR for several $p_{\text{joint}}$ thresholds.

        \begin{deluxetable}{lcccc}
        \tablecaption{Excess in the joint probability selection.
        \label{tab: joint numbers}}
        \tablehead{
        \colhead{Selection} &
        \colhead{$N_{\rm ontime}$} &
        \colhead{$\alpha N_{\rm shifted}$} &
        \colhead{Contamination Fraction} &
        \colhead{FAR ($\mathrm{yr^{-1}}$)}
        }
        \startdata
        $p_{\rm joint}>0.5$ & $2445\pm138$ & $259\pm133$ & $0.106\pm0.036$ & $29.27\pm15.10$ \\
        $p_{\rm joint}>0.75$ & $1896\pm76$ & $90\pm18$ & $0.047\pm0.007$ & $10.16\pm2.04$ \\
        $p_{\rm joint}>0.9$ & $1686\pm75$ & $59\pm11$ & $0.035\pm0.006$ & $6.68\pm1.30$ \\
        \enddata
        \tablecomments{
        Values outside the grid used to calculate the joint probability, $\rho^2_{\text{gbm}}\leq 150$ and $\rho^2_{\text{bat}}\leq 60$, are extrapolated by using the value of the nearest grid point. Errors are $1\sigma$ from the bootstrap distribution described in Appendix \ref{app: robustness}.}
        \end{deluxetable}

        Using the joint SNR distribution, we estimate the increase in sensitive volume relative to our GBM-only detection pipeline. For this estimate, we assume Euclidean geometry, which is an appropriate approximation for local-universe transients such as the binary neutron-star merger counterpart GRB170817A. Because a trigger's SNR is proportional to the event flux, which scales as $1/d^2$ for luminosity distance $d$, the sensitive volume scales roughly as $V \sim \rho^{-3/2}$. Consequently, the fractional increase in sensitive volume is:
        \begin{equation}
            \frac{\Delta V}{V} = \left(\frac{\rho^2}{\rho_{\text{th}}^2}\right)^{-3/4} - 1
        \end{equation}
        where $\rho_{\text{th}}$ is the independent detection threshold (i.e., the SNR corresponding to a fixed $p_{\text{astro}}$). This estimate neglects the relative orientations of the spacecraft, their field of view, and the energy-dependent and angle-dependent detector responses. It therefore represents an upper limit on the achievable increase in sensitive volume, intended to illustrate the potential gains from a network of detectors in which optimal alignment can be achieved more often than in a two-detector only configuration. 

        Table \ref{tab: volume increase} details the upper limit volume increase for the GRB and SGR classes defined by \cite{perera_expanding_2026}, comparing the joint search to the independent search. We calculate the new GBM threshold required to maintain a fixed $p_{\rm joint} = 0.9$ using two representative $\rho_{\rm BAT}^2$ values. While the joint search increases the sensitive volume by up to $60\%$ for certain classes, others show no improvement. As discussed in \cite{perera_expanding_2026}, the threshold depends strongly on burst parameters; classes that inherently possess a low GBM-only detection threshold (e.g., SGR3) naturally benefit less from the joint analysis, given that we did not split the joint analysis over our defined classes. 


        \begin{deluxetable}{ l c cc}
        \tablecaption{Sensitive volume increase from a joint search for $p_{\rm joint} = 0.9$.}
        \label{tab: volume increase}
        \tablehead{
          \colhead{Class} & \colhead{$\rho_{\rm GBM}^{2}$} & \multicolumn{2}{c}{$\Delta V / V$} \\
          \colhead{} & \colhead{} & \colhead{$\rho_{\rm BAT}^2 = 20$} & \colhead{$\rho_{\rm BAT}^2 = 25$}
        }
        \startdata
              GRB1         & $103$ & $29\%$ & $60\%$ \\
              GRB2         & $54$ & \nodata & \nodata \\
              GRB3         & $86$ & $13\%$ & $40\%$ \\
              GRB4         & $60$ & \nodata & $7\%$ \\
              GRB5         & $90$ & $16\%$ & $45\%$ \\
              GRB6         & $95$ & $22\%$ & $51\%$ \\
              GRB7         & $96$ & $22\%$ & $52\%$ \\
              SGR1         & $72$ & \nodata & $23\%$ \\
              SGR2         & $61$ & \nodata & $8\%$ \\
              SGR3         & $52$ & \nodata & \nodata \\
              SGR4         & $53$ & \nodata & \nodata \\
              (GRB/SGR)1   & $99$ & $25\%$ & $55\%$ \\
              (GRB/SGR)2   & $98$ & $25\%$ & $55\%$ \\
              (GRB/SGR)3   & $128$ & $52\%$ & $89\%$ \\
              (GRB/TGF)1   & $63$ & \nodata & $11\%$ \\
              (GRB/TGF)2   & $53$ & \nodata & \nodata \\
        \enddata
        \tablecomments{The column $\rho_{\rm GBM}^{2}$ is the GBM-only SNR$^2$ threshold at $p_{\rm astro} = 0.9$ for each class, taken from \cite{perera_expanding_2026}. The joint-search GBM thresholds at $p_{\rm joint} = 0.9$ are $\rho_{\rm GBM}^{2} = 73$ for $\rho_{\rm BAT}^2 = 20$ and $\rho_{\rm GBM}^{2} = 55$ for $\rho_{\rm BAT}^2 = 25$. \nodata{} indicates no volume improvement over the independent search.
        Classes (GRB1--GRB7, SGR1--SGR4, and the ambiguous GRB/SGR and GRB/TGF classes) follow the classification scheme of \cite{perera_expanding_2026}.}
        \end{deluxetable}

\section{Summary and outlook} \label{sec: summary}
    In this paper, we presented a joint detection framework for high-energy transients observed by multiple missions, and applied it to coincident Fermi/GBM and Swift/BAT triggers over the 2013-2025 data set. The framework combines the independent trigger information from the two instruments and assigns each candidate a joint probability of being a genuine astrophysical transient observed by both missions, expressed as a function of the individual detector SNRs. We showed that even a two-instrument network can improve the sensitivity to faint transients, and that under optimal conditions, the accessible detection volume can be increased by up to $\sim 60\%$. Although the foreground trigger distribution contains some contamination from background events, we quantified this contribution and found that it affects only $\sim 4\%$ of events with $p_{\text{joint}}>0.9$. 
    
    These results motivate the publication and dissemination of very low-significance sub-threshold trigger information across gamma-ray missions. More broadly, they highlight the scientific value of making time-tagged event data, or similarly detailed data products, available for joint network analyses. Such data sharing would enable searches below the nominal sensitivity threshold of individual instruments, improving the collective sensitivity to faint gamma-ray and hard X-ray transients and opening a relatively unexplored region of high-sensitivity transient phase space.
    
    By increasing the overall sensitivity to gamma-ray transients, this collaborative approach can support searches for faint counterparts to binary neutron star mergers, high-energy neutrino events, magnetar bursts, and other transient phenomena. In cases where no significant associations are found, the same framework can be used to place more stringent constraints on the rates of such events.
    
    Several natural extensions of this work remain. The framework can be expanded to include additional missions, as well as additional event properties such as localization consistency, duration, spectral hardness, and detector response information. In the present analysis, however, most events were available only through rate data, which provide limited spectral information and relatively poor localization constraints due to the four-quadrant spatial binning of the Swift/BAT data. For this reason, we restricted the analysis to the integrated rates data, with the goal of establishing a practical foundation for more comprehensive future studies.
    
    The additional information available in richer data products could also be used to construct dedicated vetoes. Such filtering would help reduce the contamination of the foreground sample by background events, thereby improving the reliability of the joint probabilities assigned to candidate transients. Finally, the analysis presented here is statistically incoherent rather than coherent: it combines the individual detection statistics from each instrument, rather than constructing a single detection statistic directly from the combined event data. Future coherent analyses that jointly model the data from multiple missions may further improve sensitivity and maximize the scientific return of high-energy transient searches across the full mission network.

\section*{Acknowledgments}
    AP thanks Oryna Ivashtenko, Edan Rein, and Dotan Gazith for helpful discussions, and to Jimmy Delaunay for sharing valuable information about the Swift/BAT instrument and data processing.
    This research was supported by grant no 2022136 from the United States - Israel Binational Science Foundation
    (BSF), Jerusalem, Israel.
    BZ is supported by a research grant from the Willner Family Leadership Institute for the Weizmann Institute of Science. 
    TV additionally acknowledges support from NSF grants 2012086 and 2309360, the Alfred P. Sloan Foundation through grant number FG-2023-20470, and the Hellman Family Faculty Fellowship during the time this work was done.

    \appendix

    \twocolumngrid

    \section{Derivation of the joint probability in terms of the measured distributions} \label{appendix: pastro}
    Here we derive the expression used to compute the probability that a
    coincident pair of triggers is truly associated, in terms of the measured
    ontime and shifted distributions. For notational convenience, let
    $x = (\rho^2_{\text{gbm}}, \rho^2_{\text{bat}})$.
    
    For every GBM trigger, the BAT follow-up analysis evaluates the detection
    statistic in the temporally coincident window, producing the ontime sample
    of $N_1$ pairs, and in time-shifted windows, producing the shifted sample of
    $N_0$ pairs. Each ontime pair is either a genuine association ($s$; both
    triggers arise from the same astrophysical transient) or an accidental
    coincidence ($n$), with prior probability $f \equiv p(s)$ and
    $p(n) = 1 - f$. The ontime sample is therefore a two-component mixture,
    \begin{equation} \label{eq: mixture}
        p(x | \mathcal{H}_1)
        = f\, p(x | s) + (1 - f)\, p(x | n),
    \end{equation}
    while the shifted sample, which by construction contains no true
    associations, samples the accidental-coincidence density,
    \begin{equation}
        p(x | n) = p(x |\mathcal{H}_0).
    \end{equation}
    
    Applying Bayes' theorem to a single ontime pair and using
    Eq. (\ref{eq: mixture}) to eliminate the (unknown) signal density
    $p(x | s)$,
    \begin{align}
        p(s | x)
        &= \frac{f\, p(x | s)}{p(x | \mathcal{H}_1)} \nonumber \\
        &= \frac{p(x | \mathcal{H}_1) - (1 - f)\, p(x | \mathcal{H}_0)}
                {p(x | \mathcal{H}_1)}.
    \end{align}

    The densities are estimated from the measured number densities of the two
    samples, $p(x | \mathcal{H}_1) \simeq n_{\rm ontime}(x)/N_1$ and
    $p(x | \mathcal{H}_0) \simeq n_{\rm shifted}(x)/N_0$, which gives
    \begin{equation} \label{eq: pjoint appendix}
        p(s | x)
        = \frac{n_{\rm ontime}(x) - \alpha\, n_{\rm shifted}(x)}
               {n_{\rm ontime}(x)},
    \end{equation}
    where $\alpha \equiv (1 - f)\, \frac{N_1}{N_0}$. The combination $\alpha\, n_{\rm shifted}(x)$ is the expected density of
    accidental coincidences within the ontime sample; integrated over the
    plane it equals $(1-f) N_1$, the expected number of background pairs among
    the ontime triggers. This provides a direct way to estimate $\alpha$
    without knowing $f$: in any region $R$ of the plane that is free of true
    associations, all ontime pairs are accidental, so
    \begin{equation} \label{eq: alpha estimate}
        \alpha =
        \frac{\int_R n_{\rm ontime}(x)\, dx}{\int_R n_{\rm shifted}(x)\, dx},
    \end{equation}
    i.e., the ratio of ontime to shifted counts in a background-dominated
    region, which is the procedure adopted in Section \ref{sec: method}.

\section{Robustness tests and error estimation}\label{app: robustness}
\begin{figure}
    \centering
    \includegraphics[width=1.0\linewidth]{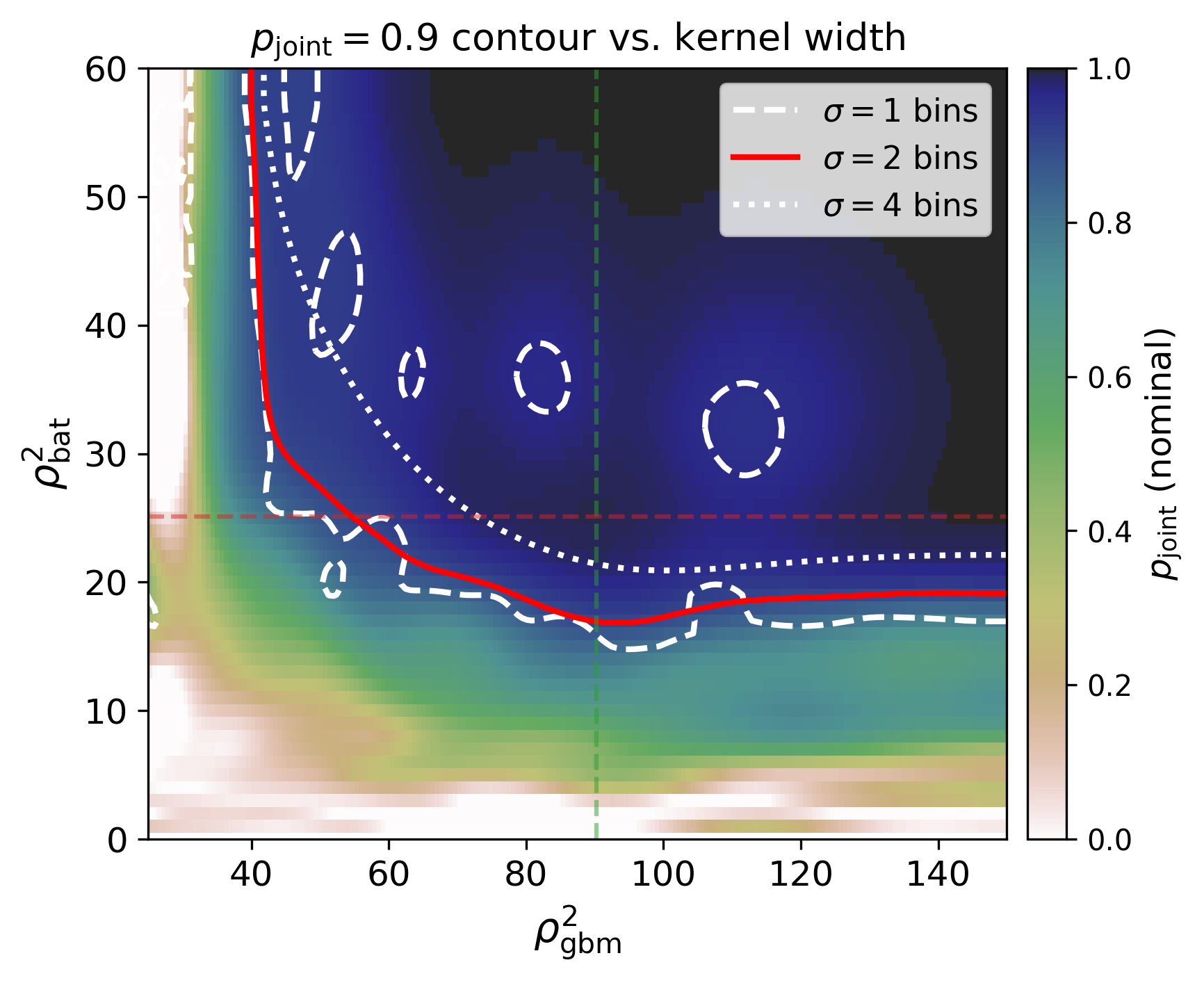}
    \caption{\textbf{Varying the smoothing kernel.} The figure shows the resulting $p_{\text{joint}}=0.9$ contours for Gaussian kernel widths of 1, 2, and 4 bins. The 2-bin kernel is the fiducial choice used in the main analysis.}
    \label{fig:kernel_varying_contours}
    \end{figure}

    To test the smoothing procedure, we varied the width of the Gaussian kernel used to smooth the distributions shown in Fig.~\ref{fig: 2d hists} and recalculated the number of triggers above the corresponding $p_{\text{joint}}=0.9$ threshold. Figure~\ref{fig:kernel_varying_contours} shows the resulting $p_{\text{joint}}=0.9$ contours for the different kernel widths. Doubling the kernel width produces a smoother contour, but shifts it toward higher SNR in both detectors, resulting in fewer triggers above $p_{\text{joint}}=0.9$. In contrast, halving the kernel width produces a more jagged contour without substantially changing the number of selected triggers. Specifically, reducing the Gaussian kernel width from 2 to 1 bins changes $N(p_{\text{joint}}>0.9)$ from 1652 to 1651, a decrease of only 0.1\%, while increasing the kernel width to 4 bins reduces the number to 1466, a decrease of 11.3\%.
    
    In addition, we performed a bootstrap resampling of the triggers in the ontime and shifted distributions. For each realization, triggers were sampled with replacement and the full analysis described in the main text was repeated. We generated 1000 bootstrap realizations and used the resulting distributions to estimate the $1\sigma$ uncertainties reported in Table~\ref{tab: joint numbers}.

    \onecolumngrid

    \section{Tables}

    \startlongtable
    \begin{deluxetable*}{llccccccccc}
    \tabletypesize{\scriptsize}
    \tablecaption{Swift/BAT catalog GRBs associated with triggers of the joint GBM--BAT search over 2013--2025.\label{appendix: bat recovered grbs}}
    \tablehead{
    \colhead{GRB} & \colhead{Trigger Time} & \colhead{$\rho_{\rm gbm}^{2}$} & \colhead{$\rho_{\rm bat}^{2}$} & \colhead{$p_{\rm joint}$} & \colhead{BAT $T_{90}$} & \colhead{Pipe Dur.} & \colhead{GBM $T_{90}$} & \colhead{BAT R.A., Decl.} & \colhead{Pipe R.A., Decl.} & \colhead{Offset} \\
    \colhead{} & \colhead{(UTC)} & \colhead{} & \colhead{} & \colhead{} & \colhead{(s)} & \colhead{(s)} & \colhead{(s)} & \colhead{(deg)} & \colhead{(deg)} & \colhead{(deg)}
    }
    \startdata
    GRB130305A\tablenotemark{a} & 2013-03-05 11:39:19.05 & 61.6 & 19.8 & 0.856 & 52.80 & 0.030 & 25.60 & 116.8, 52.0 & 134.9, 56.5 & 11.4 \\
    GRB130327A\tablenotemark{a} & 2013-03-27 01:47:33.75 & 92.2 & 17.6 & 0.932 & 6.37 & 6.573 & \nodata & 92.0, 55.7 & 77.6, 58.1 & 8.2 \\
    GRB130420B & 2013-04-20 12:56:34.09 & 221.4 & 1178.5 & 1.000 & 12.64 & 4.869 & 13.82 & 183.1, 54.4 & 196.7, 51.6 & 8.6 \\
    GRB130502A & 2013-05-02 17:50:31.94 & 329.6 & 247.0 & 1.000 & 7.57 & 3.606 & 3.33 & 138.6, -0.1 & 141.1, 2.2 & 3.5 \\
    GRB130515A & 2013-05-15 01:21:17.90 & 1835.0 & 4228.9 & 1.000 & 0.30 & 0.133 & 0.26 & 283.4, -54.3 & 290.2, -50.0 & 5.9 \\
    GRB130528A & 2013-05-28 16:41:28.71 & 1404.9 & 165.1 & 1.000 & 640.00 & 6.573 & 55.55 & 139.4, 87.3 & 72.5, 79.6 & 9.7 \\
    GRB130609A & 2013-06-09 03:05:10.13 & 417.1 & 747.4 & 1.000 & 7.06 & 2.671 & 5.38 & 152.7, 24.1 & 150.0, 27.6 & 4.2 \\
    GRB130609B\tablenotemark{a} & 2013-06-09 21:41:50.39 & 48.3 & 3.6 & 0.082 & 211.17 & 2.671 & 191.49 & 53.8, -40.2 & 41.4, -21.6 & 21.3 \\
    GRB130610A & 2013-06-10 03:12:16.05 & 427.4 & 1.2 & 0.029 & 47.72 & 6.573 & 21.76 & 224.4, 28.2 & 226.2, 26.4 & 2.4 \\
    GRB130626A & 2013-06-26 10:51:03.83 & 180.4 & 185.7 & 1.000 & 0.16 & 0.179 & 1.73 & 273.1, -9.5 & 265.4, 0.2 & 12.4 \\
    GRB130708A & 2013-07-08 11:43:09.44 & 474.0 & 133.8 & 1.000 & 11.20 & 6.573 & 14.08 & 17.5, -0.0 & 17.9, 3.4 & 3.4 \\
    GRB130716A & 2013-07-16 10:36:53.67 & 474.7 & 710.0 & 1.000 & 87.67 & 0.596 & 0.77 & 179.5, 63.1 & 175.4, 70.4 & 7.5 \\
    GRB130803A & 2013-08-03 10:02:54.68 & 1584.2 & 234.4 & 1.000 & 43.57 & 2.671 & 7.62 & 220.3, -2.5 & 221.5, -6.8 & 4.5 \\
    GRB130806A & 2013-08-06 02:51:33.82 & 104.0 & 443.7 & 0.999 & 5.78 & 6.573 & \nodata & 35.9, 67.5 & 359.3, 63.1 & 15.7 \\
    GRB130912A & 2013-09-12 08:34:58.56 & 602.6 & 361.6 & 1.000 & 0.28 & 1.466 & 0.51 & 47.6, 14.0 & 44.3, 9.1 & 5.9 \\
    GRB131004A & 2013-10-04 21:41:04.10 & 473.3 & 425.3 & 1.000 & 1.54 & 1.466 & 1.15 & 296.1, -3.0 & 298.9, -7.5 & 5.3 \\
    GRB131024B\tablenotemark{a} & 2013-10-24 21:35:28.12 & 45.7 & 32.5 & 0.920 & 96.82 & 0.133 & 45.31 & 144.5, 44.3 & 145.8, 42.7 & 1.8 \\
    GRB131031A & 2013-10-31 11:33:34.65 & 1474.1 & 1650.5 & 1.000 & 6.40 & 2.671 & 7.42 & 29.6, -1.6 & 34.7, -5.5 & 6.4 \\
    GRB131117A & 2013-11-17 00:34:08.91 & 79.3 & 168.1 & 0.998 & 10.88 & 2.671 & \nodata & 332.4, -31.8 & 356.4, -27.8 & 21.2 \\
    GRB131128A & 2013-11-28 15:06:25.26 & 381.3 & 523.0 & 1.000 & 3.00 & 1.979 & 1.98 & 355.3, 31.3 & 2.8, 27.1 & 7.7 \\
    GRB131224B & 2013-12-24 03:25:09.13 & 131.6 & 121.6 & 1.000 & 8.35 & 2.671 & \nodata & 163.7, -14.2 & 159.4, -18.1 & 5.7 \\
    GRB140102A & 2014-01-02 21:17:38.57 & 49.9 & 180.8 & 0.928 & 55.08 & 0.030 & 3.65 & 211.9, 1.3 & 208.2, 3.0 & 4.1 \\
    GRB140108A & 2014-01-08 17:18:49.54 & 284.5 & 23.6 & 0.995 & 95.23 & 1.086 & 91.39 & 325.1, 58.7 & 319.7, 60.9 & 3.5 \\
    GRB140301A\tablenotemark{a} & 2014-03-01 15:24:50.98 & 63.0 & 36.8 & 0.962 & 27.80 & 2.671 & \nodata & 69.5, -34.2 & 60.0, -34.4 & 7.9 \\
    GRB140304A & 2014-03-04 13:22:37.26 & 221.2 & 513.8 & 1.000 & 14.78 & 6.573 & 31.23 & 30.6, 33.5 & 33.2, 31.5 & 3.0 \\
    GRB140320A & 2014-03-20 02:12:46.02 & 297.9 & 348.4 & 1.000 & 0.51 & 0.327 & 2.30 & 281.8, -11.2 & 279.7, -5.7 & 5.9 \\
    GRB140402A & 2014-04-02 00:10:06.97 & 351.4 & 468.2 & 1.000 & 0.90 & 0.098 & 0.32 & 207.6, 6.0 & 200.9, 7.9 & 6.9 \\
    GRB140408A & 2014-04-08 13:15:56.65 & 224.8 & 1.0 & 0.029 & 3.63 & 6.573 & 7.68 & 290.7, -12.6 & 286.6, -14.2 & 4.4 \\
    GRB140506A & 2014-05-06 21:07:37.38 & 1506.4 & 206.0 & 1.000 & 111.10 & 1.086 & 64.13 & 276.8, -55.6 & 275.4, -54.1 & 1.7 \\
    GRB140512A & 2014-05-12 19:31:49.17 & 179.9 & 210.0 & 1.000 & 154.11 & 1.979 & 147.97 & 289.4, -15.1 & 291.3, -20.3 & 5.5 \\
    GRB140515A & 2014-05-15 09:12:34.21 & 41.2 & 111.3 & 0.912 & 23.42 & 2.671 & \nodata & 186.1, 15.1 & 181.5, 21.0 & 7.4 \\
    GRB140516A\tablenotemark{a} & 2014-05-16 20:30:54.84 & 56.9 & 82.9 & 0.970 & 0.26 & 0.133 & \nodata & 253.0, 40.0 & 259.9, 28.9 & 12.5 \\
    GRB140703A & 2014-07-03 00:37:17.34 & 258.9 & 51.1 & 1.000 & 68.64 & 6.573 & 83.97 & 13.0, 45.1 & 14.6, 44.0 & 1.6 \\
    GRB140706A\tablenotemark{a} & 2014-07-06 19:33:32.82 & 54.3 & 62.9 & 0.951 & 47.84 & 2.671 & 43.78 & 49.3, -38.1 & 56.6, -38.1 & 5.7 \\
    GRB140716A-1 & 2014-07-16 10:27:58.09 & 1793.7 & 154.1 & 1.000 & 124.80 & 1.086 & 168.25 & 108.1, -60.1 & 106.4, -63.2 & 3.2 \\
    GRB140817A & 2014-08-17 07:02:02.49 & 1615.8 & 7093.0 & 1.000 & 258.34 & 4.869 & 16.13 & 127.2, 58.2 & 133.8, 50.0 & 9.0 \\
    GRB140818B\tablenotemark{a} & 2014-08-18 18:44:16.56 & 145.7 & 94.5 & 1.000 & 31.55 & 2.671 & 20.99 & 271.2, -1.4 & 275.6, -6.3 & 6.6 \\
    GRB141005A & 2014-10-05 05:13:08.09 & 2768.3 & 142.3 & 1.000 & 3.47 & 2.671 & 3.39 & 291.1, 36.1 & 301.3, 41.9 & 9.8 \\
    GRB141102A & 2014-11-02 12:51:40.81 & 1745.1 & 408.3 & 1.000 & 14.40 & 0.441 & 2.62 & 208.6, -47.1 & 212.9, -50.8 & 4.7 \\
    GRB141205A & 2014-12-05 08:05:17.73 & 401.8 & 388.2 & 1.000 & 1.66 & 0.596 & 1.28 & 92.9, 37.9 & 95.1, 26.6 & 11.4 \\
    GRB141229A & 2014-12-29 11:49:00.37 & 3102.8 & 1395.5 & 1.000 & 6.40 & 1.086 & 13.82 & 72.5, -19.3 & 79.4, -16.4 & 7.2 \\
    GRB150101A & 2015-01-01 06:28:53.73 & 120.5 & 224.7 & 0.999 & 0.06 & 0.022 & 0.48 & 312.6, 36.7 & 309.8, 36.8 & 2.2 \\
    GRB150101B & 2015-01-01 15:23:34.46 & 408.8 & 146.2 & 1.000 & 0.01 & 0.030 & 0.08 & 188.0, -11.0 & 183.9, -8.1 & 5.0 \\
    GRB150110B & 2015-01-10 22:08:31.19 & 900.7 & 490.3 & 1.000 & 10.58 & 6.573 & 2.56 & 289.4, 32.5 & 289.6, 37.1 & 4.6 \\
    GRB150120A & 2015-01-20 02:57:46.87 & 212.8 & 753.7 & 1.000 & 1.20 & 1.086 & 3.33 & 10.3, 34.0 & 358.3, 26.8 & 12.6 \\
    GRB150202A & 2015-02-02 23:10:04.87 & 67.8 & 154.5 & 0.997 & 25.47 & 2.671 & \nodata & 39.2, -33.1 & 18.9, -12.8 & 27.5 \\
    GRB150204A & 2015-02-04 06:31:08.45 & 552.2 & 548.6 & 1.000 & 14.00 & 6.573 & 11.01 & 160.3, -64.0 & 148.2, -61.5 & 6.1 \\
    GRB150309A\tablenotemark{a} & 2015-03-09 23:04:11.80 & 73.3 & 1.2 & 0.000 & 242.04 & 6.573 & 52.48 & 277.0, 86.4 & 125.2, 64.9 & 28.3 \\
    GRB150323C\tablenotemark{a} & 2015-03-23 17:05:07.27 & 80.5 & 31.6 & 0.972 & 159.66 & 4.869 & 43.26 & 192.6, 50.2 & 196.4, 52.3 & 3.2 \\
    GRB150530A & 2015-05-30 11:42:18.36 & 4535.1 & 137.8 & 1.000 & 6.90 & 6.573 & 7.17 & 327.5, 57.5 & 329.1, 58.8 & 1.5 \\
    GRB150607A & 2015-06-07 07:55:11.88 & 177.5 & 411.1 & 1.000 & 25.99 & 4.869 & 26.17 & 140.0, 68.4 & 124.6, 66.1 & 6.4 \\
    GRB150711A\tablenotemark{a} & 2015-07-11 18:23:05.65 & 60.7 & 57.1 & 0.985 & 70.96 & 1.086 & 82.18 & 221.6, -35.5 & 214.7, -38.7 & 6.4 \\
    GRB150817A & 2015-08-17 02:05:25.15 & 990.4 & 421.0 & 1.000 & 38.00 & 0.804 & 35.58 & 249.6, -12.1 & 256.3, -21.6 & 11.5 \\
    GRB151006A\tablenotemark{a} & 2015-10-06 09:55:06.50 & 49.6 & 38.2 & 0.930 & 211.06 & 2.671 & 93.44 & 147.4, 70.5 & 156.1, 72.7 & 3.5 \\
    GRB151027B\tablenotemark{a} & 2015-10-27 22:41:24.62 & 62.0 & 58.7 & 0.989 & 80.00 & 6.573 & \nodata & 76.2, -6.4 & 75.5, -4.4 & 2.2 \\
    GRB151029A & 2015-10-29 07:49:40.22 & 262.0 & 166.1 & 1.000 & 8.95 & 4.869 & \nodata & 38.5, -35.4 & 33.7, -30.8 & 6.1 \\
    GRB151114A & 2015-11-14 09:59:34.55 & 63.4 & 202.4 & 0.992 & 4.86 & 1.086 & \nodata & 120.9, -61.0 & 115.4, -68.4 & 7.7 \\
    GRB151122A & 2015-11-22 17:00:46.44 & 136.1 & 141.7 & 1.000 & 36.80 & 6.573 & 51.20 & 299.7, -19.9 & 303.0, -22.2 & 3.9 \\
    GRB151205A\tablenotemark{a} & 2015-12-05 15:46:02.54 & 55.7 & 39.0 & 0.940 & 64.00 & 2.671 & 56.32 & 229.3, 35.8 & 246.6, 14.8 & 26.1 \\
    GRB151229A & 2015-12-29 06:50:28.45 & 3178.5 & 1038.9 & 1.000 & 1.44 & 1.466 & 3.46 & 329.4, -20.7 & 329.2, -21.4 & 0.7 \\
    GRB160117B & 2016-01-17 13:59:30.23 & 165.6 & 321.7 & 1.000 & 11.54 & 1.086 & \nodata & 132.2, -16.3 & 130.8, -12.3 & 4.3 \\
    GRB160127A\tablenotemark{a} & 2016-01-27 08:43:11.19 & 71.5 & 15.2 & 0.692 & 6.16 & 1.086 & \nodata & 226.0, 0.1 & 221.5, 3.8 & 5.8 \\
    GRB160327A & 2016-03-27 09:16:12.04 & 120.6 & 1618.4 & 0.999 & 33.74 & 6.573 & \nodata & 146.7, 54.0 & 163.5, 61.5 & 11.6 \\
    GRB160408A & 2016-04-08 06:25:43.96 & 2465.7 & 1253.1 & 1.000 & 0.32 & 0.327 & 1.06 & 122.6, 71.1 & 115.9, 69.0 & 3.1 \\
    GRB160417A & 2016-04-17 04:23:44.33 & 61.3 & 104.0 & 0.987 & 14.55 & 4.869 & \nodata & 120.3, 7.6 & 117.5, -0.9 & 9.0 \\
    GRB160519A & 2016-05-19 00:17:34.37 & 188.4 & 1.2 & 0.029 & 35.62 & 1.979 & 98.56 & 71.1, 31.2 & 77.7, 28.8 & 6.2 \\
    GRB160612A & 2016-06-12 20:12:47.60 & 2866.3 & 6006.6 & 1.000 & 0.25 & 0.179 & 0.29 & 348.4, -25.4 & 353.2, -32.4 & 8.2 \\
    GRB160716A & 2016-07-16 01:08:33.09 & 300.8 & 72.5 & 1.000 & 6.39 & 6.573 & \nodata & 190.5, -61.4 & 206.4, -55.6 & 10.1 \\
    GRB160815A & 2016-08-15 11:45:13.48 & 1018.7 & 1352.4 & 1.000 & 8.57 & 4.869 & 7.10 & 288.6, 84.3 & 301.7, 85.6 & 1.7 \\
    GRB160819A & 2016-08-19 20:28:04.18 & 185.6 & 234.7 & 1.000 & 67.20 & 1.086 & 33.54 & 114.1, -22.3 & 119.0, -21.1 & 4.7 \\
    GRB160821B & 2016-08-21 22:29:13.34 & 287.6 & 1129.4 & 1.000 & 0.48 & 0.098 & 1.09 & 280.0, 62.4 & 277.5, 58.5 & 4.1 \\
    GRB160917A & 2016-09-17 11:30:19.75 & 162.0 & 124.0 & 1.000 & 14.52 & 1.466 & 19.46 & 295.7, 46.4 & 295.2, 51.9 & 5.5 \\
    GRB161017A & 2016-10-17 17:52:10.64 & 296.4 & 1331.8 & 1.000 & 217.05 & 6.573 & 37.89 & 142.8, 43.1 & 145.5, 43.0 & 2.0 \\
    GRB161129A & 2016-11-29 07:12:12.40 & 313.5 & 48.6 & 1.000 & 35.54 & 2.671 & 36.10 & 316.2, 32.1 & 308.6, 29.5 & 7.0 \\
    GRB170113A & 2017-01-13 10:04:08.21 & 74.9 & 246.4 & 0.998 & 20.30 & 6.573 & 49.15 & 61.7, -71.9 & 77.1, -70.9 & 5.0 \\
    GRB170126A & 2017-01-26 11:30:44.90 & 5784.7 & 77.5 & 1.000 & 9.54 & 6.573 & 13.82 & 263.6, -64.8 & 260.7, -67.4 & 2.9 \\
    GRB170208A & 2017-02-08 18:11:18.61 & 1333.1 & 175.8 & 1.000 & 7.45 & 4.869 & 7.17 & 166.5, -46.8 & 169.5, -44.5 & 3.1 \\
    GRB170318B & 2017-03-18 15:27:52.85 & 142.2 & 219.3 & 1.000 & 1.07 & 1.086 & 4.10 & 284.3, 6.3 & 291.2, 18.0 & 13.5 \\
    GRB170516A\tablenotemark{a} & 2017-05-16 12:49:22.50 & 59.2 & 39.8 & 0.952 & 36.77 & 4.869 & \nodata & 41.5, -55.9 & 36.2, -54.4 & 3.4 \\
    GRB170705A & 2017-07-05 02:46:00.75 & 1957.2 & 3687.7 & 1.000 & 223.20 & 2.671 & 22.78 & 191.7, 18.3 & 189.8, 19.3 & 2.1 \\
    GRB170710B & 2017-07-10 08:10:15.06 & 166.6 & 138.1 & 1.000 & 54.40 & 6.573 & 42.24 & 43.1, 42.7 & 36.0, 56.7 & 14.8 \\
    GRB170728A\tablenotemark{a} & 2017-07-28 06:53:29.62 & 68.4 & 78.9 & 0.998 & 1.25 & 1.466 & \nodata & 58.9, 12.2 & 55.3, 23.3 & 11.7 \\
    GRB170906A & 2017-09-06 00:44:39.20 & 186.3 & 0.1 & 0.029 & 88.11 & 6.573 & 78.85 & 203.9, -47.1 & 213.4, -43.3 & 7.7 \\
    GRB170906B & 2017-09-06 00:55:46.57 & 290.5 & 128.5 & 1.000 & 19.20 & 1.466 & 11.52 & 232.2, -28.3 & 233.1, -26.7 & 1.7 \\
    GRB170912B & 2017-09-12 06:33:50.28 & 371.8 & 466.2 & 1.000 & 17.88 & 2.671 & 13.57 & 215.5, -62.0 & 214.0, -63.2 & 1.4 \\
    GRB171007A & 2017-10-07 11:57:39.12 & 203.1 & 153.2 & 1.000 & 68.42 & 1.979 & 3.46 & 135.5, 42.8 & 149.2, 60.5 & 19.5 \\
    GRB171010B\tablenotemark{a} & 2017-10-10 20:59:25.80 & 41.0 & 34.6 & 0.884 & 30.40 & 1.979 & 25.34 & 34.1, -54.4 & 42.5, -50.7 & 6.3 \\
    GRB171120A & 2017-11-20 13:20:42.01 & 3481.4 & 1626.3 & 1.000 & 64.00 & 2.671 & 44.06 & 163.8, 22.5 & 165.3, 25.9 & 3.7 \\
    GRB180102A & 2018-01-02 15:49:45.49 & 199.9 & 338.3 & 1.000 & 11.56 & 6.573 & 13.31 & 203.1, 62.2 & 202.4, 63.9 & 1.7 \\
    GRB180205A & 2018-02-05 04:25:29.45 & 108.6 & 208.5 & 0.999 & 15.54 & 0.596 & 15.36 & 126.8, 11.5 & 134.5, 9.6 & 7.8 \\
    GRB180620B\tablenotemark{a} & 2018-06-20 15:50:37.46 & 61.5 & 11.5 & 0.507 & 223.97 & 1.979 & 46.72 & 357.5, -58.0 & 353.5, -48.2 & 10.1 \\
    GRB180706A & 2018-07-06 08:25:09.35 & 154.0 & 306.6 & 1.000 & 42.44 & 2.671 & 38.14 & 181.7, 66.0 & 169.7, 66.7 & 4.8 \\
    GRB180718A & 2018-07-18 01:57:44.60 & 325.4 & 112.3 & 1.000 & 0.08 & 0.133 & 0.08 & 336.0, 2.8 & 331.7, 23.8 & 21.5 \\
    GRB180720C & 2018-07-20 22:24:03.54 & 374.4 & 106.6 & 1.000 & 122.99 & 6.573 & 23.30 & 265.7, -26.6 & 266.2, -26.9 & 0.6 \\
    GRB180805A & 2018-08-05 09:04:49.50 & 66.4 & 105.6 & 0.996 & 1.68 & 1.086 & \nodata & 167.6, -45.3 & 199.2, -45.2 & 22.1 \\
    GRB180805B & 2018-08-05 13:02:36.60 & 864.6 & 721.2 & 1.000 & 122.24 & 0.327 & 0.96 & 25.9, -17.5 & 25.8, -1.4 & 16.1 \\
    GRB180812A & 2018-08-12 08:22:31.78 & 154.4 & 31.6 & 0.998 & 16.53 & 3.606 & 38.40 & 245.8, 74.7 & 209.9, 74.2 & 9.5 \\
    GRB181010A & 2018-10-10 05:55:59.67 & 184.4 & 232.7 & 1.000 & 15.56 & 0.804 & 9.73 & 52.6, -23.0 & 57.6, -18.4 & 6.6 \\
    GRB181126A & 2018-11-26 09:54:09.98 & 399.0 & 1083.1 & 1.000 & 2.09 & 0.327 & 2.11 & 152.3, -29.7 & 149.3, -29.3 & 2.7 \\
    GRB190109B & 2019-01-09 11:56:09.66 & 130.8 & 191.2 & 1.000 & 6.51 & 1.086 & 8.19 & 55.6, 63.6 & 70.8, 58.5 & 8.9 \\
    GRB190326A & 2019-03-26 07:35:28.92 & 1647.5 & 6.5 & 0.229 & 0.08 & 0.022 & 55.80 & 341.7, 39.9 & 351.8, 47.8 & 10.7 \\
    GRB190511A\tablenotemark{a} & 2019-05-11 07:14:48.97 & 75.9 & 50.1 & 0.994 & 27.66 & 0.098 & 27.65 & 126.4, -20.3 & 123.9, -25.8 & 6.0 \\
    GRB190611A\tablenotemark{a} & 2019-06-11 17:49:02.60 & 47.0 & 4.1 & 0.074 & 41.76 & 2.671 & \nodata & 324.7, -56.1 & 327.7, -54.5 & 2.3 \\
    GRB190821A\tablenotemark{a} & 2019-08-21 17:10:34.34 & 55.3 & 20.4 & 0.832 & 57.10 & 1.466 & 59.90 & 250.1, -34.0 & 259.1, -41.3 & 10.2 \\
    GRB190828B & 2019-08-28 13:00:09.73 & 58.8 & 124.8 & 0.977 & 715.86 & 6.573 & 63.49 & 251.8, 27.3 & 249.5, 23.6 & 4.3 \\
    GRB191011A & 2019-10-11 04:35:56.80 & 204.1 & 1426.5 & 1.000 & 7.35 & 3.606 & 25.09 & 44.7, -27.9 & 45.7, -31.7 & 3.9 \\
    GRB191101A\tablenotemark{a} & 2019-11-01 21:08:06.15 & 57.6 & 15.5 & 0.716 & 142.71 & 2.671 & \nodata & 251.9, 43.7 & 256.3, 32.4 & 11.9 \\
    GRB200109A\tablenotemark{a} & 2020-01-09 01:45:51.84 & 44.0 & 19.0 & 0.602 & 112.00 & 1.086 & 40.96 & 307.1, 53.0 & 296.2, 54.9 & 6.7 \\
    GRB200219A & 2020-02-19 07:36:49.25 & 4533.9 & 15014.5 & 1.000 & 288.00 & 0.327 & 1.15 & 342.6, -59.1 & 345.1, -46.5 & 12.7 \\
    GRB200227A & 2020-02-27 07:20:19.36 & 630.6 & 5533.6 & 1.000 & 30.33 & 6.573 & 24.32 & 56.4, 9.5 & 56.6, 3.4 & 6.1 \\
    GRB200228B\tablenotemark{a} & 2020-02-28 11:14:43.38 & 112.1 & 39.0 & 0.968 & 7.37 & 2.671 & 14.08 & 252.0, 17.0 & 252.5, 6.1 & 10.8 \\
    GRB200409A & 2020-04-09 03:20:00.20 & 113.8 & 223.5 & 0.999 & 16.00 & 0.804 & \nodata & 60.6, -50.2 & 53.5, -43.1 & 8.6 \\
    GRB200411A & 2020-04-11 04:29:02.56 & 508.8 & 1324.8 & 1.000 & 0.22 & 0.179 & 1.44 & 47.7, -52.3 & 64.2, -54.9 & 10.1 \\
    GRB200416A & 2020-04-16 07:05:19.25 & 1236.2 & 581.9 & 1.000 & 6.05 & 6.573 & 9.34 & 335.7, -7.5 & 332.9, -2.4 & 5.8 \\
    GRB200509A & 2020-05-09 07:00:56.32 & 97.6 & 721.3 & 0.999 & 828.23 & 6.573 & \nodata & 116.4, -4.6 & 115.9, -3.2 & 1.5 \\
    GRB200630A & 2020-06-30 01:49:36.00 & 442.6 & 1633.3 & 1.000 & 14.84 & 4.869 & 7.17 & 91.4, -60.8 & 98.6, -60.3 & 3.6 \\
    GRB200711A & 2020-07-11 11:04:36.81 & 857.4 & 1529.5 & 1.000 & 29.36 & 2.671 & 29.44 & 286.0, -0.1 & 291.3, 4.5 & 7.1 \\
    GRB200716C & 2020-07-16 22:57:43.50 & 7077.4 & 5517.2 & 1.000 & 86.57 & 4.869 & 5.31 & 196.0, 29.6 & 188.5, 34.3 & 7.9 \\
    GRB200729A\tablenotemark{a} & 2020-07-29 19:38:10.46 & 80.3 & 71.5 & 0.998 & 122.00 & 6.573 & \nodata & 184.4, 45.6 & 190.7, 39.6 & 7.6 \\
    GRB200809B & 2020-08-09 15:41:27.42 & 775.5 & 527.9 & 1.000 & 4.20 & 0.804 & 17.41 & 15.9, -73.8 & 15.8, -68.4 & 5.4 \\
    GRB201020A & 2020-10-20 05:47:44.19 & 1442.9 & 259.7 & 1.000 & 14.36 & 6.573 & 21.50 & 261.2, 31.4 & 259.3, 34.6 & 3.5 \\
    GRB201223A & 2020-12-23 17:58:28.63 & 490.5 & 355.5 & 1.000 & 29.00 & 6.573 & 33.28 & 132.7, 71.2 & 120.9, 59.1 & 13.0 \\
    GRB210104B & 2021-01-04 21:10:21.21 & 60.6 & 707.3 & 0.985 & 19.94 & 6.573 & \nodata & 53.6, 37.9 & 53.2, 36.6 & 1.4 \\
    GRB210222B & 2021-02-22 22:37:25.71 & 119.2 & 109.1 & 0.999 & 12.11 & 6.573 & \nodata & 154.6, -14.9 & 160.6, -8.7 & 8.5 \\
    GRB210226A & 2021-02-26 04:43:57.57 & 597.8 & 741.8 & 1.000 & 20.80 & 2.671 & 18.88 & 124.2, 57.6 & 123.2, 53.2 & 4.4 \\
    GRB210305A & 2021-03-05 19:03:05.42 & 129.4 & 111.4 & 1.000 & 67.52 & 6.573 & 61.70 & 319.8, 34.5 & 320.4, 34.8 & 0.6 \\
    GRB210308A & 2021-03-08 06:37:59.96 & 23231.6 & 2555.7 & 1.000 & 6.19 & 3.606 & 5.44 & 67.1, 37.4 & 60.0, 39.0 & 5.8 \\
    GRB210610A & 2021-06-10 15:03:42.69 & 735.3 & 1017.8 & 1.000 & 13.62 & 6.573 & 8.19 & 204.3, 14.5 & 203.8, 9.8 & 4.7 \\
    GRB210723A & 2021-07-23 14:46:09.87 & 250.2 & 28.0 & 0.999 & 55.06 & 2.671 & 42.75 & 121.7, -32.9 & 116.5, -33.6 & 4.4 \\
    GRB210725A & 2021-07-25 03:47:05.90 & 325.7 & 93.6 & 1.000 & 53.54 & 6.573 & 47.10 & 215.4, -1.2 & 216.2, 4.8 & 6.0 \\
    GRB210725B & 2021-07-25 12:00:50.21 & 43.1 & 110.6 & 0.920 & 417.91 & 3.606 & \nodata & 192.9, 17.1 & 187.0, 9.0 & 10.0 \\
    GRB210730A & 2021-07-30 04:57:30.47 & 4581.0 & 1581.2 & 1.000 & 3.86 & 2.671 & 4.16 & 149.6, 69.7 & 146.8, 70.0 & 1.0 \\
    GRB210905A & 2021-09-05 00:12:40.08 & 70.9 & 241.4 & 0.998 & 778.40 & 6.573 & \nodata & 309.1, -44.4 & 327.3, -38.5 & 14.8 \\
    GRB210930A & 2021-09-30 02:47:01.75 & 188.7 & 523.0 & 1.000 & 11.81 & 6.573 & \nodata & 197.4, 48.6 & 184.8, 44.3 & 9.7 \\
    GRB211129A\tablenotemark{a} & 2021-11-29 09:51:34.69 & 51.0 & 21.3 & 0.805 & 643.69 & 4.869 & 111.11 & 274.6, 31.8 & 283.5, 34.8 & 8.1 \\
    GRB211207A\tablenotemark{a} & 2021-12-07 20:52:57.95 & 80.6 & 74.8 & 0.998 & 3.73 & 1.979 & \nodata & 149.6, -24.4 & 157.0, -11.4 & 14.7 \\
    GRB220408A & 2022-04-08 05:46:06.12 & 1162.8 & 226.3 & 1.000 & 17.24 & 4.869 & 17.92 & 202.4, 47.1 & 203.3, 46.0 & 1.2 \\
    GRB220521A & 2022-05-21 23:20:21.89 & 280.3 & 12.0 & 0.705 & 13.55 & 1.466 & 13.57 & 275.2, 10.4 & 276.2, 9.9 & 1.1 \\
    GRB220826A & 2022-08-26 11:55:26.03 & 484.8 & 2580.0 & 1.000 & 11.14 & 6.573 & 16.64 & 206.4, -44.0 & 206.7, -39.4 & 4.7 \\
    GRB220907A & 2022-09-07 14:05:26.81 & 259.7 & 45.7 & 0.999 & 8.88 & 1.466 & 8.70 & 268.9, -20.3 & 267.5, -20.5 & 1.3 \\
    GRB221216A & 2022-12-16 11:24:05.55 & 159.5 & 178.5 & 1.000 & 228.53 & 6.573 & 184.32 & 326.1, -34.4 & 321.0, -16.1 & 18.9 \\
    GRB230216A\tablenotemark{a} & 2023-02-16 14:48:35.53 & 96.0 & 40.2 & 0.987 & 91.20 & 1.466 & \nodata & 114.0, -8.0 & 120.6, -18.4 & 12.3 \\
    GRB230322B & 2023-03-22 21:10:40.03 & 319.9 & 301.1 & 1.000 & 11.80 & 4.869 & \nodata & 16.6, -47.7 & 22.3, -50.5 & 4.7 \\
    GRB230903A & 2023-09-03 17:22:57.09 & 97.6 & 105.2 & 0.999 & 2.54 & 1.086 & 2.82 & 9.9, -40.9 & 353.5, -23.9 & 21.8 \\
    GRB231118A & 2023-11-18 17:16:32.47 & 5552.2 & 8440.3 & 1.000 & 151.49 & 4.869 & 5.76 & 4.8, -48.0 & 3.4, -51.5 & 3.5 \\
    GRB231230A & 2023-12-30 01:29:12.55 & 356.3 & 408.9 & 1.000 & 15.24 & 6.573 & 17.15 & 245.2, 58.1 & 239.8, 55.9 & 3.7 \\
    GRB241025A\tablenotemark{a} & 2024-10-25 01:37:16.30 & 90.3 & 20.8 & 0.977 & 130.09 & 0.098 & 123.91 & 333.7, 83.6 & 214.2, 80.7 & 13.6 \\
    GRB241026A & 2024-10-26 22:42:32.51 & 195.7 & 357.6 & 1.000 & 24.98 & 1.086 & 16.38 & 293.4, 58.0 & 306.0, 53.6 & 8.3 \\
    GRB241030A\tablenotemark{a} & 2024-10-30 05:48:13.49 & 114.9 & 12.4 & 0.719 & 173.43 & 4.869 & 165.63 & 343.2, 80.4 & 324.7, 76.3 & 5.5 \\
    GRB241030B & 2024-10-30 18:34:22.39 & 4810.1 & 999.5 & 1.000 & 6.62 & 4.869 & 6.85 & 50.8, 34.4 & 45.7, 38.0 & 5.4 \\
    GRB241115A & 2024-11-15 13:18:26.30 & 3769.3 & 15.0 & 0.681 & 3.70 & 3.606 & 3.07 & 86.8, -0.7 & 85.0, 6.4 & 7.3 \\
    GRB250101A\tablenotemark{a} & 2025-01-01 13:22:52.18 & 42.2 & 24.3 & 0.760 & 34.22 & 3.606 & \nodata & 37.1, 19.2 & 46.8, -8.3 & 29.1 \\
    GRB250128B & 2025-01-28 16:22:53.76 & 108.0 & 837.7 & 0.999 & 0.48 & 0.098 & 1.12 & 231.4, -0.5 & 250.6, 20.9 & 28.5 \\
    GRB250321A & 2025-03-21 00:42:40.64 & 997.2 & 138.4 & 1.000 & 6.21 & 3.606 & 6.40 & 295.1, 21.1 & 298.9, 25.2 & 5.4 \\
    GRB250509A & 2025-05-09 22:33:31.34 & 141.6 & 121.2 & 1.000 & 54.33 & 2.671 & 69.12 & 46.5, -38.9 & 44.4, -29.8 & 9.2 \\
    \enddata
    \tablenotetext{a}{Within the axis ranges of Fig.~\ref{fig: pjoint} ($\rho^2_{\rm gbm} \leq 150$, $\rho^2_{\rm bat} \leq 100$).}
    \tablecomments{A pipeline trigger is associated with a BAT catalog GRB when its trigger time falls within the interval $[t_{\rm BAT} - 1\,{\rm s},\, t_{\rm BAT} + T_{90} + 1\,{\rm s}]$ and its GBM MLE localization lies within $30^\circ$ of the BAT ground position. When several triggers match the same GRB, the highest-$p_{\rm joint}$ entry is kept. $\rho_{\rm gbm}^2$ and $\rho_{\rm bat}^2$ are the pipeline GBM and BAT matched-filter SNRs squared; BAT $T_{90}$, Pipe $\Delta t$, and GBM $T_{90}$ are the Swift duration, pipeline search timescale, and Fermi/GBM catalog duration (\nodata if no GBM catalog association); Offset is the angular separation between the Swift ground position and the pipeline MLE localization. Of 155 unique GRBs, 42 have BAT $T_{90} < 6.573\,$s (within the longest search timescale) and 113 are longer.}
    \end{deluxetable*}

    \section{Rates and sky localization plots} \label{appendix: gbm bat plots}
    
        \begin{figure*}
            \centering
            \includegraphics[width=0.45\linewidth]{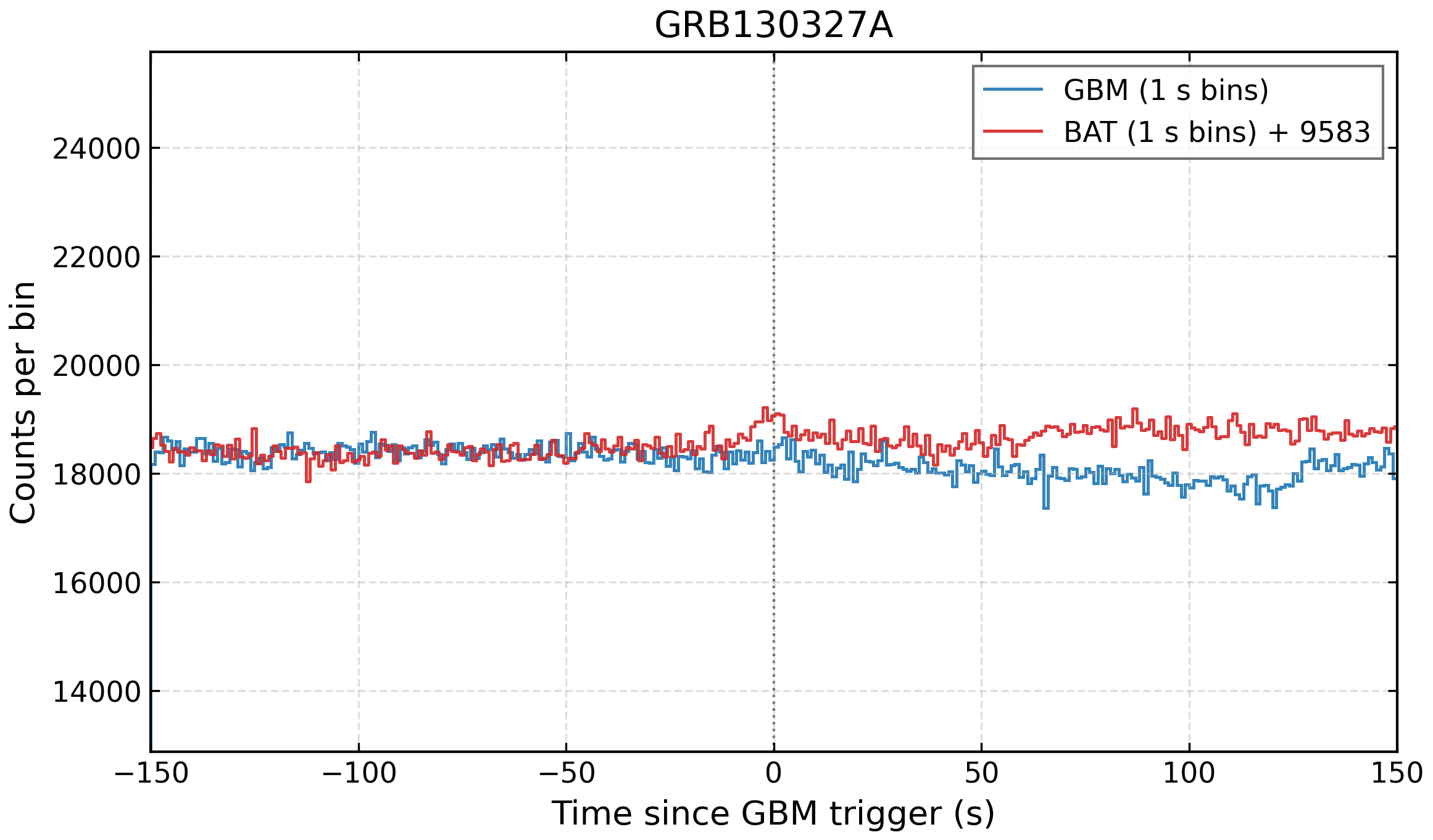}
            \includegraphics[width=0.5\linewidth]{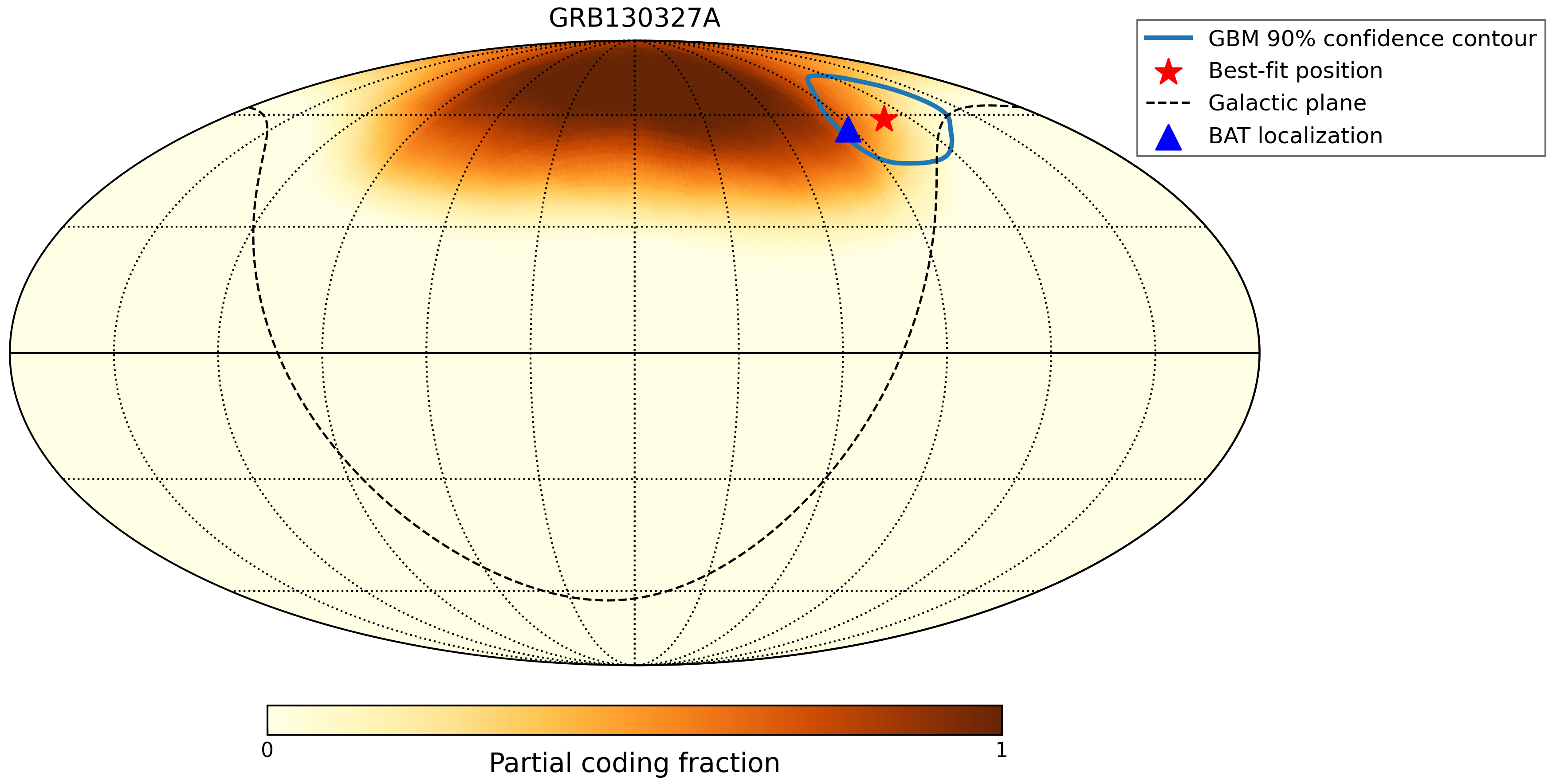}
            \includegraphics[width=0.45\linewidth]{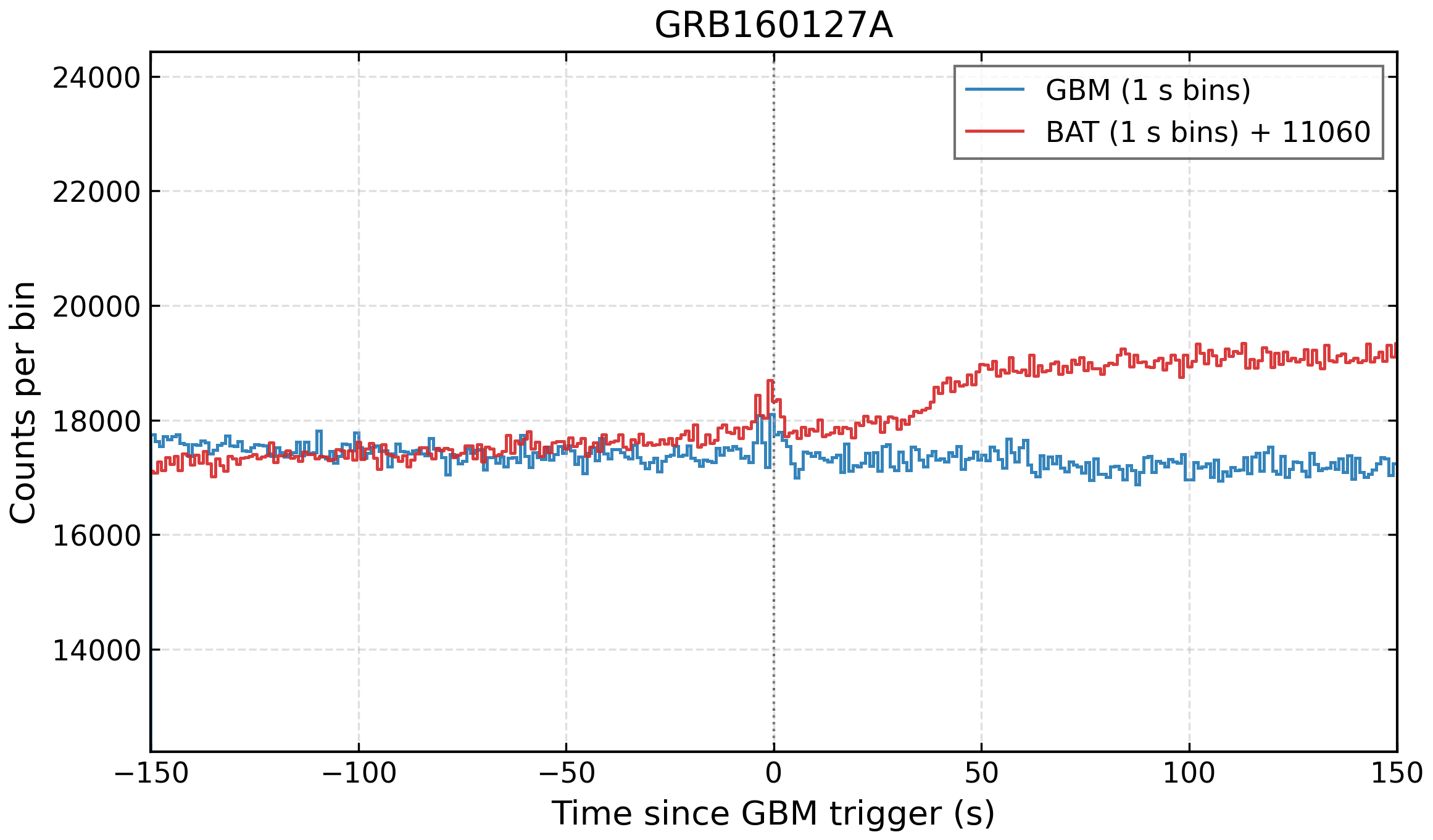}
            \includegraphics[width=0.5\linewidth]{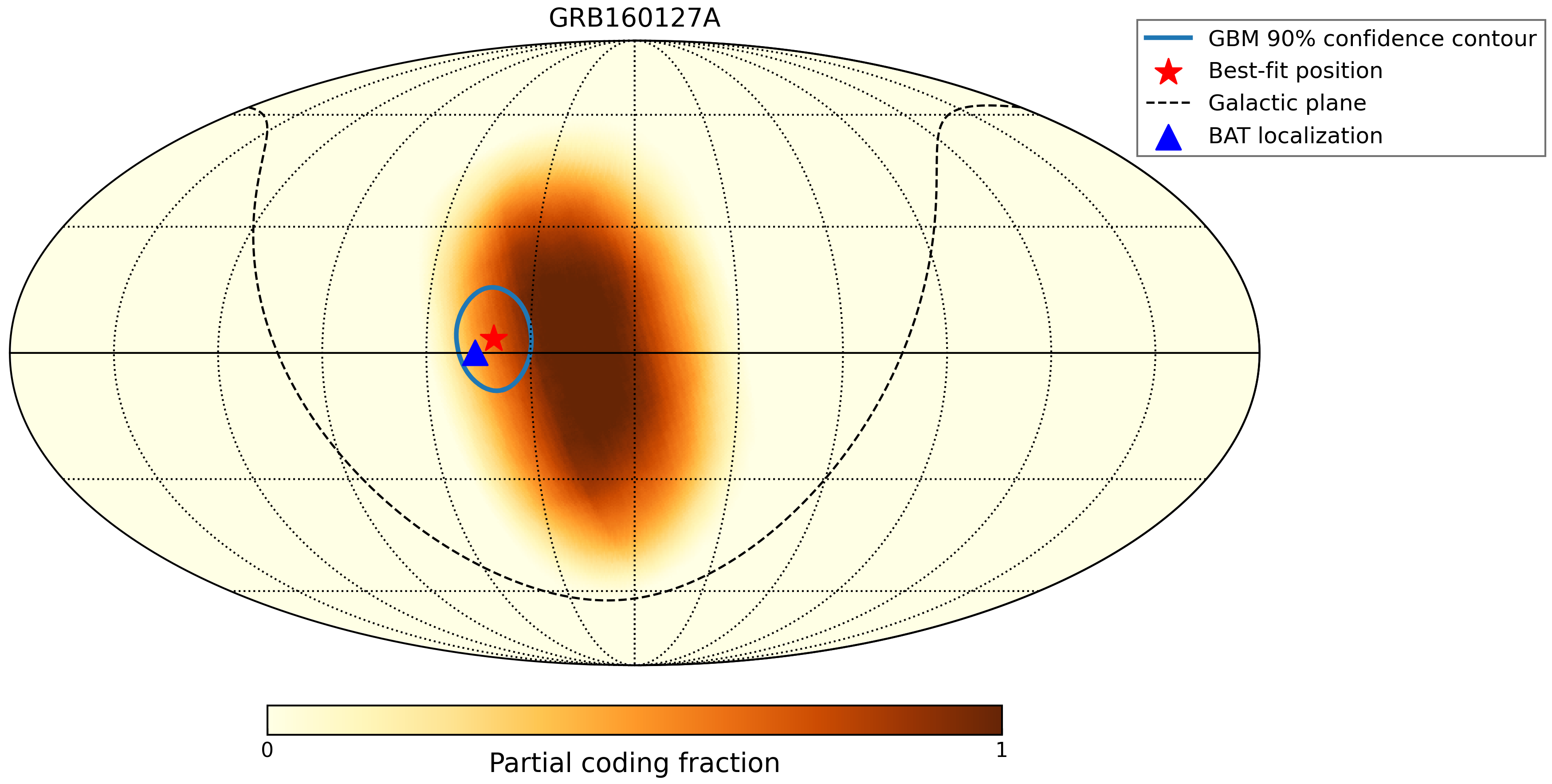}
            \includegraphics[width=0.45\linewidth]{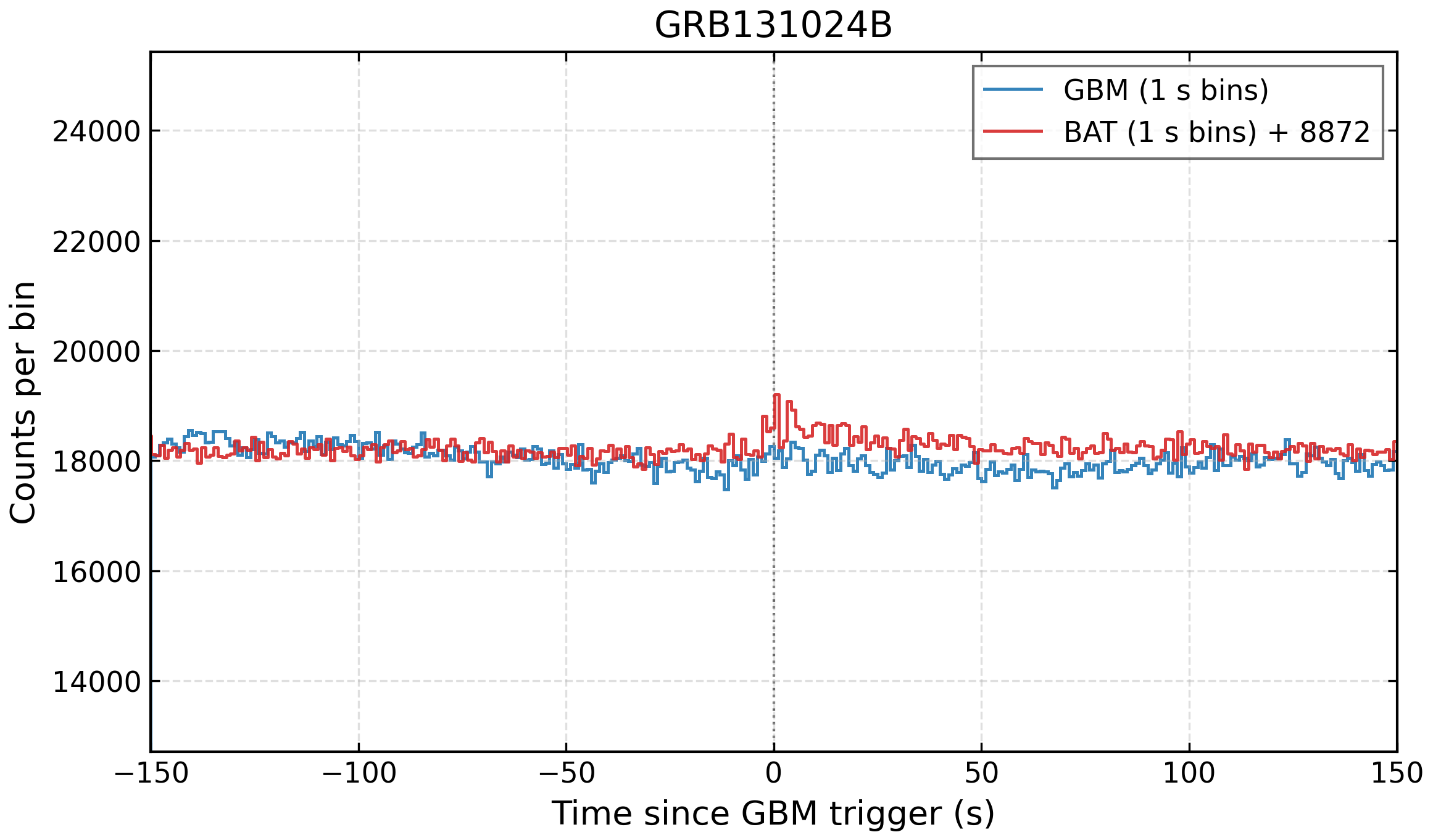}
            \includegraphics[width=0.5\linewidth]{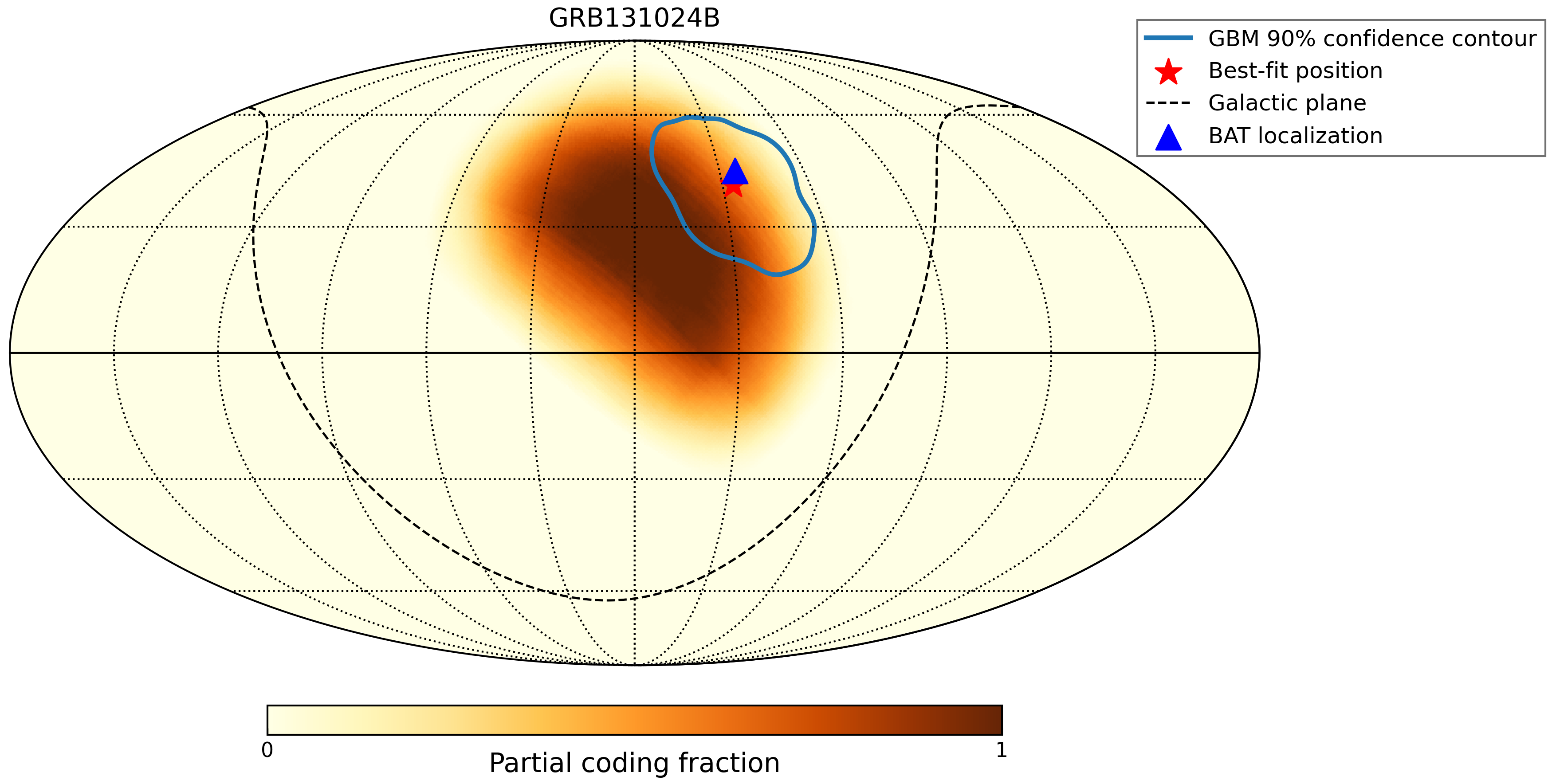}
            \includegraphics[width=0.45\linewidth]{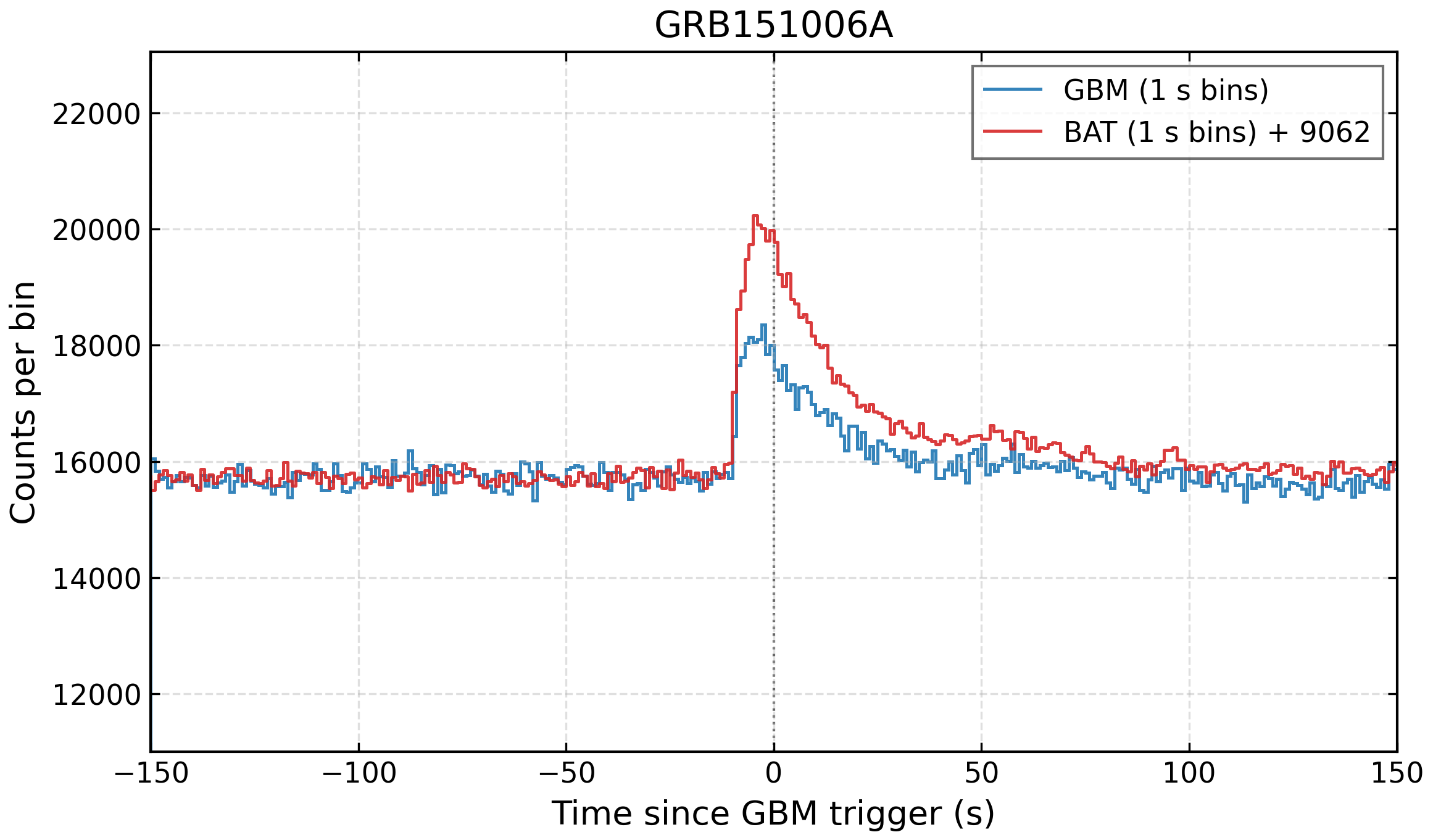}
            \includegraphics[width=0.5\linewidth]{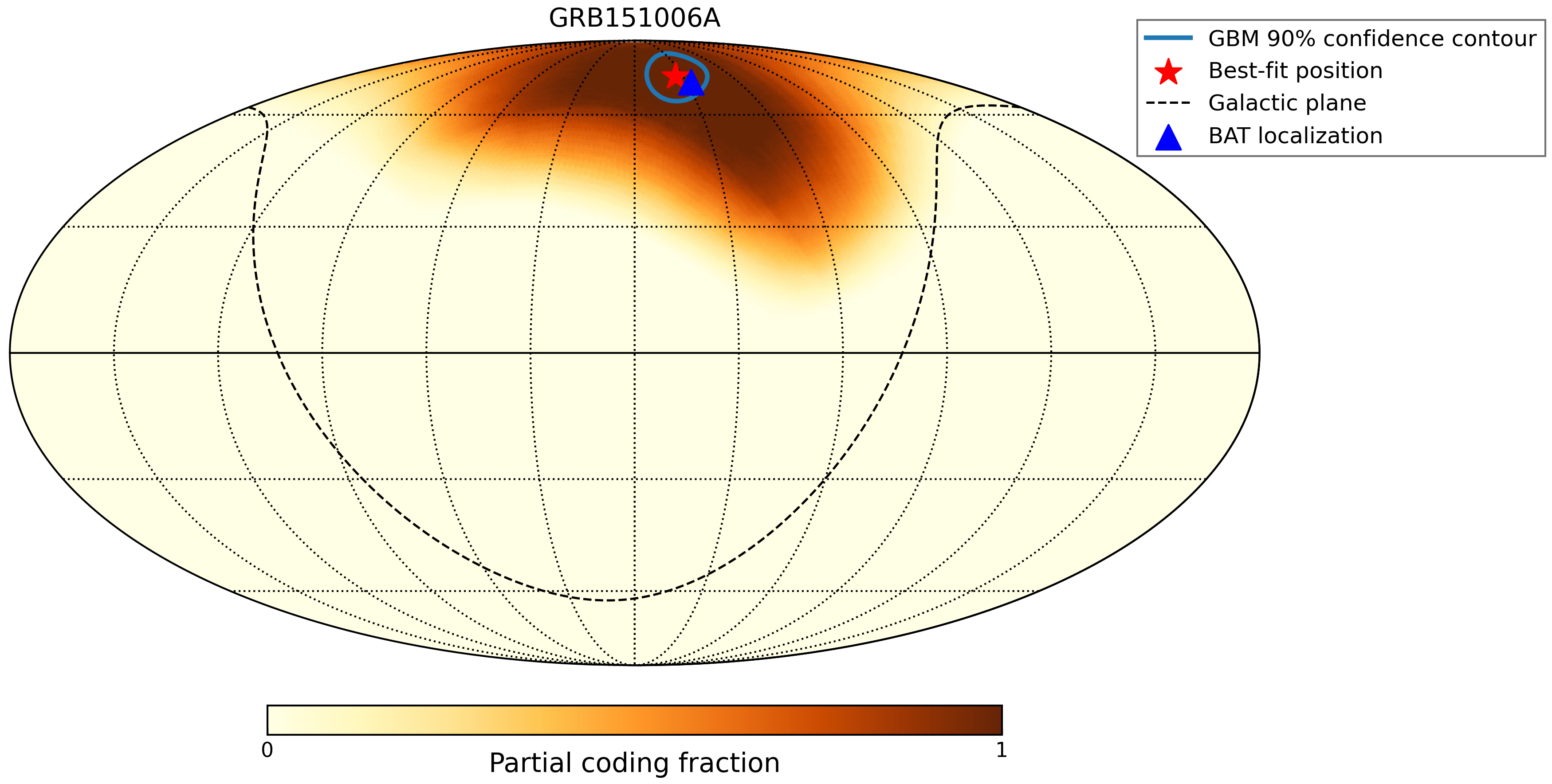}
            \caption{\textbf{Rates and sky localizations of BAT catalog GRBs recovered by the joint search.}
            \textit{Left:} BAT rate data and GBM photon counts summed over their respective energy channels. The BAT detection statistic is computed from this summed light curve, whereas the GBM detection pipeline employs a two-dimensional matched filter using spectral templates across the energy channels, which are summed here for illustration only. The BAT rates have been shifted vertically to match the GBM background level for visual comparison.
            \textit{Right:} The GBM 90\% localization contour overlaid on the BAT partial-coding map, with the pipeline best-fit position (star) and BAT ground position (triangle).
            Top two rows: GRB130327A and GRB160127A that are sub-threshold in each instrument individually ($p_{\rm astro}=0.82$, $p_{\rm bat}=0.69$ and $p_{\rm astro}=0.58$, $p_{\rm bat}=0.63$) but are promoted by the joint statistic to $p_{\rm joint}=0.93$ and $0.69$, respectively. The BAT positions lie within the GBM localization contours, with angular offsets of $8.2^\circ$ and $5.8^\circ$.
            Bottom two rows: GRB131024B and GRB151006A, long catalog bursts ($T_{90}=97$~s and $211$~s) recovered through short emission peaks at the $0.133$~s and $2.671$~s templates ($\rho^2_{\rm gbm}=45.7$ and $49.6$), yet assigned $p_{\rm joint}>0.9$.}
            \label{fig: catalog validation}
        \end{figure*}

\bibliography{references}{}
\bibliographystyle{aasjournalv7}

\end{document}